\documentclass[preprint,aps,nofootinbib,showpacs]{revtex4}
\usepackage{graphicx}
\usepackage{epsfig}
\usepackage{bm}
\usepackage{amssymb}

\newcommand{\be}{\begin{eqnarray}}
\newcommand{\ee}{\end{eqnarray}}

\begin{document}

\def\be{\begin{equation}}
\def\ee{\end{equation}}
\def\Be{\begin{eqnarray}}
\def\Ee{\end{eqnarray}}
\def\ba{\begin{array}}
\def\ea{\end{array}}

\title{Neutrino induced reactions related to the $\nu$-process nucleosynthesis of
${}^{92}$Nb and ${}^{98}$Tc}
\author{Myung-Ki Cheoun$^{1)}$ \footnote{Corresponding author :
cheoun@ssu.ac.kr}, Eunja Ha$^{1)}$, T. Hayakawa$^{2)}$, Satoshi
Chiba$^{2)}$, Ko Nakamura$^{3)}$ Toshitaka Kajino$^{3,4)}$, Grant
J. Mathews$^{5)}$}
\address{
1)Department of Physics, Soongsil University, Seoul 156-743, Korea
\\
2)Advanced Science Research Center, Japan Atomic Energy Agency,
2-4 Shirakata-shirane, Tokai, Ibaraki 319-1195, Japan \\
3)National Astronomical Observatory, Mitaka, Tokyo 181-8589,
Japan \\
4)Department of Astronomy, Graduate School of Science, University
of Tokyo, 7-3-1 Hongo, Tokyo 113-0033, Japan \\
5)Center for Astrophysics, Department of Physics, University of
Notre Dame, IN 46556, USA}

\date{\today}

\begin{abstract}
It has recently been proposed that ${}^{92}_{41}$Nb and
${}^{98}_{43}$Tc may have been formed in the $\nu$-process. We investigate the neutrino induced reactions related to the
$\nu$-process origin of the two odd-odd nuclei. The main neutrino reactions for ${}^{92}_{41}$Nb are
the charged-current (CC) $^{92}$Zr($\nu_e,e^{-}$)$^{92}$Nb   and
the neutral-current (NC) $^{93}$Nb(${\nu} ({\bar \nu}) ,~{\nu}^{'} ({\bar \nu})^{'}$ n
)$^{92}$Nb reactions. The main reactions for ${}^{98}_{43}$Tc, are the CC
reaction $^{98}$Mo($\nu_e,e^-$)$^{98}$Tc and the NC reaction
$^{99}$Ru(${\nu} ({\bar \nu}) ,~{\nu}^{'} ({\bar \nu})^{'}$
p)$^{98}$Tc. Our calculations are carried out
using the quasi-particle random phase approximation. Numerical
results are presented for the  energy  and temperature
dependent cross sections. Since charge exchange reactions by
neutrons may also lead to the formation of ${}^{92}_{41}$Nb and
${}^{98}_{43}$Tc, we discuss the feasibility of the
$^{92}$Mo(n,p)$^{92}$Nb and $^{98}$Ru(n,p)$^{98}$Tc reactions to produce
these nuclei.
\end{abstract}

\maketitle

\section{Introduction}
The neutrino ($\nu$) process involves $\nu$-induced reactions on various
nuclei during core collapse supernovae (SN). This process has been proposed as
the origin of some rare isotopes of light and heavy elements
\cite{Woosley90}. The cosmic abundances of these nuclei could thus be valuable tools
 for studying  neutrino spectra from
 supernovae (SNe) \cite{Yoshida05,Heger05}, and for constraining
neutrino oscillation and/or other $\nu$-physics parameters
\cite{Yoshida06}.

Among the many heavy elements, only the two isotopes  $^{138}$La and
$^{180}$Ta are currently thought to be synthesized primarily by the $\nu$
process \cite{Woosley90,Heger05}. These two isotopes have similar
features: they cannot be produced by either  ${\beta}^{+}$, EC, or ${\beta}^{-}$ decays
since in either case stable isobars shield against these decays.
Not surprisingly then, the isotopic abundance ratios, $^{138}$La/$^{139}$La and
$^{180}$Ta/$^{181}$Ta, are quite small, {\it i.e.} 0.0902\% and 0.012\%,
respectively \cite{Hayakawa10}, making them Nature's rarest isotopes.

In principle, any nuclide can be synthesized by the $\nu$ process in
SN explosions. The produced abundances, however, are usually negligibly small because
of the relevant reactions are mediated by a  weak interaction compared to production via strong or electromagnetic interactions for the  other major
nucleosynthesis processes such as the $s-$, $r-$, and $\gamma -$
processes. Thus,  the $\nu$ process
can only  play a dominant role in the case of very rare isotopes that cannot be produced by other means.
In the $\nu$ process, a nucleus can be synthesized by either a charged current (CC) or a neutral
current (NC) reaction. Previous studies have concluded that
contributions from the CC reactions are generally larger than those of the NC
reactions for heavy nuclei \cite{Heger05,Ch09-1,Ch09-2,Ch10}.

In a recent work \cite{Haya11}, it has been pointed out that the nuclear chart
around $^{92}$Nb and $^{98}$Tc is quite similar to that of
$^{138}$La and $^{180}$Ta as shown in Figs. 1 and 2. Although both nuclei
are unstable, their half-lives, 3.47$\times 10^7$ yr
for ${}^{92}$Nb and 4.2$\times 10^6$ yr for ${}^{98}$Tc, are long
enough to be observed on stellar surfaces or to be incorporated into meteorites. Moreover, they are shielded from $\beta^{+}$, EC, or $\beta^{-}$
decays because of the presence of neighboring stable isobars
\cite{Yin00,Munker00,Yin02,Meyer03,Clayton77}.

The isotopic abundance ratio  of ${}^{92}$Nb/${}^{93}$Nb is
known to be $\sim10^{-3} - 10^{-5}$ at the time of solar system formation
\cite{Maria02,Maria05}. This is comparable to the isotopic rations for
$^{138}$La/$^{139}$La and $^{180}$Ta/$^{181}$Ta.
Therefore, it has been proposed \cite{Haya11} that the two nuclei  $^{92}$Nb and $^{98}$Tc may have a $\nu-$process origin.

The main $\nu$-process reactions for $^{92}$Nb are the
$^{92}$Zr($\nu_e,e^-$)$^{92}$Nb CC reaction and the $^{93}$Nb(${\nu} ( {\bar
\nu} ),~{\nu}' ({\bar \nu}^{'} )$ n )$^{92}$Nb NC reaction, whereby
 neutrino-induced neutron emission to $^{92}$Nb is followed by a
NC neutrino reaction. For ${}^{98}$Tc, the CC reaction,
$^{98}$Mo($\nu_e$,e$^-$)$^{98}$Tc, and the NC reaction,
$^{99}$Ru(${\nu} ({\bar \nu}),~{\nu} ({\bar \nu}^{'}  )~ p)
{}^{98}$Tc, with neutrino-induced proton emission  to $^{98}$Tc are believed to be
 the main reactions. Another NC reaction, $^{99}$Tc(${\nu} ({\bar
\nu}  ),~{\nu}' ({\bar \nu}^{'}  )~ n) {}^{98}$Tc, might also be
 possible because $^{99}$Mo can easily $\beta$ decay to
 $^{99}$Tc. However, the  half life of $^{98}$Tc is 4.2$\times 10^6$ yr. This is longer than
 that of $^{99}$Tc, 2.11$\times 10^5$ yr, so that $^{98}$Tc may be difficult to produce by this NC
 reaction.

Moreover, if $^{98}$Tc is produced by $\nu$-induced reactions,
it might $\beta$ decay to $^{98}$Ru$^*$ which  subsequently decays
to its ground state with the emission of  0.74536 and 0.65243 MeV $\gamma$
rays by E2 transitions. This situation closely resembles $^{26}$Al,
whose life time is 7.4$\times 10^7$ yr and decays to $^{26}$Mg
with a 1.809 $\gamma$ ray (E2 transition) as observed by the COMPTEL detector
on the Compton Gamma-Ray Observatory (CGRO) \cite{Nasa3}.

We presume a two step process in the $\nu$-induced reaction: the
1st step is the formation of excited nuclei by incident $\nu$'s; and
the 2nd is the decay process to other ground states with some
particle emission. To describe the 2nd decay process, one needs to
consider the branching ratios for the  decay processes by using
a Hauser-Feshbach (HF) statistical model
\cite{Haus52,Naka05,Suzuki06,Yoshida08,Suzuki09}. One also needs
calculations of the transmission coefficients for the emitted particles. In
this work, we made this calculation using the method of Refs.
\cite{Naka05,Suzuki06}.

The nuclear structure of $^{92}$Nb and $^{98}$Tc are key
ingredients for this calculation. For example, the excited states
with low spins in $^{92}$Nb are strongly populated by a
Gamow-Teller (GT) transitions from the 0$^+$ ground state of the
$^{92}$Zr seed nucleus. Our scheme for describing such excited
states makes use of  the standard quasi-particle random phase
approximation (QRPA). For the NC reaction,
$^{93}$Nb(${\nu}$,~${\nu}$')$^{93}$Nb and
$^{99}$Ru(${\nu}$,~${\nu}$')$^{99}$Ru, we generate the ground and
excited states of the odd-even target nuclei, $^{93}$Nb and $^{99}$Ru,
by applying a one quasi-particle operator to the even-even  nuclei, $^{92}$Zr and
$^{98}$Ru, which are assumed to be in  the BCS ground state.

In Sec. II, we address a brief summary of our QRPA framework used
in $\nu$-induced reactions. In Sec. III, numerical
cross sections for neutrino induced reactions on relevant nuclei
are given  in terms of the incident neutrino energy. Their temperature
dependence is also presented for astrophysical applications under
the assumption of a Fermi Dirac distribution for the SN neutrinos.
A discussion about the roles of charge exchange reactions by neutron
capture on nuclei is also added to the results. A summary and
conclusions are presented in Sec. IV.

\section{Theoretical framework}

Our QRPA formalism for the $\nu ({\bar \nu})$-nucleus ($\nu
({\bar \nu}) - A)$ reaction has been detailed in our previous papers
\cite{Ch09-1,Ch09-2,Ch10}. Results from the QRPA have successfully
described relevant $\nu$-induced reaction data for $^{12}$C,
$^{56}$Fe, $^{56}$Ni, $^{138}$La and $^{180}$Ta as well as
$\beta$, 2$\nu 2 \beta$ and 0$\nu$2$\beta$ decays. In particular,
 double beta $(2 \beta)$ decay is well known to be sensitive to
the nuclear structure and has more data than the $\nu$-induced
reaction data. Therefore, it could be a useful tool  for the estimation of
$\nu$-process reaction rates.

Charge exchange reactions, A(n,p)B or B(p,n)A, also provide  viable
tests of  nuclear models and one can deduce the neutrino induced
reaction rates from these reactions because Gamow Teller (GT) transitions
account for most of the
strength in both the nucleon exchange reaction and  the neutrino reactions,
particularly in the low energy region.

Here we summarize two important characteristics regarding our calculation
compared to other
QRPA approaches. First, we utilize the Brueckner G matrix  for the
two-body interactions inside nuclei by solving the Bethe-Salpeter
equation based on the Bonn CD potential for the nucleon-nucleon
interactions in free space. This procedure reduces some of the ambiguities regarding
nucleon-nucleon interactions inside nuclei.

Secondly, we include neutron-proton (np) pairing as well as
neutron-neutron (nn) and proton-proton (pp) pairing correlations.
Consequently, both CC and NC reactions can be  described within a
single framework.

The contribution from np pairing, however, has been shown to be only of order
1 $\sim$ 2 \% for the weak interaction on $^{12}$C, such as
$\beta^{\pm}$ decay and the $\nu - ^{12} $C reaction
\cite{Ch09-1,Ch09-2}. Such a small effect is easily understood
because the energy gaps between the neutron and proton energy spaces
in light nuclei are too large to be effective. However,  in medium-heavy
nuclei, such as $^{56}$Fe and $^{56}$Ni, the np pairing effect
accounts for 20 $\sim$ 30 \% of the total cross section
\cite{Ch09-2}. Therefore, in the heavy nuclei considered in this
work, np pairing should be included.

The np pairing has two isospin contributions, T = 1 and T = 0. These
correspond to J = 0 and J = 1 pairings, respectively. Since
the J = 0 (T = 1) pairing couples a state to its
time reversed state, the  shape is almost spherical. Hence, the
J = 0 (T = 1) np pairing can be easily included in a spherical
symmetric model.

The J = 1 (T = 0) np pairing, which is partly associated with the
tensor force, however, leads to a non-spherical shape, {\it i.e.}
deformation. Therefore, in principle, the J = 1 (T = 0) np coupling cannot be
included in a  spherically symmetrically model. However,
if we use a renormalized strength constant for the np pairing,
$g_{np}$, as a parameter to be fitted to the empirical pairing
gaps, the J = 1 (T = 0) pairing can be incorporated implicitly
even in a spherical symmetric model because the fitted $g_{np}$
may include effectively the deformation of the nucleus.

The empirical np pairing gap is easily extracted from data on  mass-excesses.
The  theoretical pairing gap $\delta_{np}^{th.}$ is calculated as
the difference between the total energies with and without np pairing
correlations \cite{Ch93}
\begin{equation}
\delta_{np}^{th.} = - [ (H_0^{'} + E_1^{'} + E_2^{'}) - (H_0 + E_1
+ E_2) ],
\end{equation}
where $H_0^{'}(H_0)$ is the Hartree-Fock energy of the  ground state
with (without) np pairing and $E_1^{'} + E_2^{'} ( E_1 + E_2)$ is
the  sum of the lowest two quasi-particles energies with (without) np
pairing correlations. More detailed discussion of this is given at Ref.
\cite{Ch93}.

In our QRPA calculation, the ground state of a target nucleus is described by
the BCS vacuum for the quasi-particle which comprises nn, pp and
np pairing correlations. Excited states, $\vert m; J^{\pi} M
\rangle$, in the  compound nucleus are generated by operating the
following one phonon operator on the initial BCS state
\begin{equation} Q^{+,m}_{JM}  = {\mathop\Sigma_{k l \mu^{'}
\nu^{'} }} [ X^{m}_{(k \mu^{'} l \nu^{'} J)} C^{+}(k \mu^{'} l
\nu^{'} J M)
 - Y^{m}_{(k \mu^{'} l \nu^{'} J)} {\tilde C}(k \mu^{'} l \nu^{'}
  J M)] ~,\end{equation}
where the pair creation and annihilation operators, $C^+$ and ${\tilde
C}$, are defined as
\begin{equation}
 C^{+} (k \mu^{'} l \nu^{'} J M)  =  {\mathop\Sigma_{m_{k} m_{l}}}
C^{JM}_{j_{k} m_{k} j_{l} m_{l}} a^{+}_{l \nu^{'}} a^{+}_{k
\mu^{'}}~,~ {\tilde C}(k \mu^{'} l \nu^{'} J M)  =  (-)^{J-M} C(k
\mu^{'} l \nu^{'} J - M)~,
\end{equation}
where $a^{+}_{l \nu^{'}}$ is  a quasi-particle creation operator, and
the $C^{JM}_{j_{k} m_{k} j_{l} m_{l}}$ are Clebsh-Gordan coefficients. Here
Roman letters indicate single particle states, while Greek letters
with a prime mean quasi-particle types 1 or 2.

If the neutron-proton pairing is neglected, quasi-particles become
quasi-protons and quasi-neutrons, and the phonon operator is easily
decoupled into two different phonon operators. One is for charge
changing reactions such as nuclear $\beta$ decay and CC
neutrino reactions. The other is for charge conserving
reactions such as electromagnetic and NC neutrino reactions. The
amplitudes $X_{a {\alpha}^{'}, b {\beta}^{'}}$ and $Y_{a
{\alpha}^{'}, b {\beta}^{'}}$, which stand for forward and
backward going amplitudes from the ground states to excited states,
are obtained from the QRPA equation. A detailed derivation for this procedure was
given in Refs. \cite{Ch93,Ch09-2}

By using the phonon operator $Q^{+,m}_{JM}$ in Eq.(2), we obtain
the following expression for the CC neutrino reactions
\begin{eqnarray}
& &< QRPA || {\cal {\hat O}}_{\lambda } || ~ \omega ; JM
>  \\ \nonumber
= & & {\mathop\Sigma_{a \alpha^{'} b \beta^{'}}}  [ {\cal N}_{a
\alpha^{'} b \beta^{'} } < a \alpha^{'} || {\cal {\hat
O}}_{\lambda}  || b \beta^{'}
>  ~[ u_{pa \alpha^{'}} v_{nb
\beta^{'}} X_{a \alpha^{'} b \beta^{'}} + v_{pa \alpha^{'}} u_{nb
\beta^{'}} Y_{a \alpha^{'} b \beta^{'}} ]~,
\end{eqnarray}
where $ {\cal N}_{a \alpha^{'} b \beta^{'}} (J) \equiv  {\sqrt{ 1 -
\delta_{ab} \delta_{\alpha^{'}  \beta^{'} } (-1)^{J + T} }}/ ({1 +
\delta_{ab}\delta_{\alpha^{'}  \beta^{'} } }) $. This form is also
easily reduced to the result produced in the pn QRPA when
the np pairing correlations are not included \cite{Ring08}
\begin{equation} < QRPA || {\cal {\hat O}}_{\lambda } || ~ \omega ; JM
>  = {\mathop\Sigma_{ap bn}}  [ {\cal N}_{a p b n } < a p || {\cal
{\hat O}}_{\lambda}  || b n> ~[ u_{pa} v_{nb} X_{a p b n} + v_{pa
} u_{nb } Y_{a p b n} ]~.
\end{equation}

Since NC reactions for $^{93}$Nb and $^{98}$Tc occur in odd-even
nuclei, we need to properly describe the ground state of odd-even
nuclei. The standard QRPA treats the ground state of  even-even
nuclei as the BCS vacuum, so it is not easily applicable to reactions on odd-even nuclei.

Our formalism to deal with such NC reactions is based upon the
quasi-particle shell model (QSM) \cite{Ch10}. First, we generate
low energy spectra of odd-even nuclei by applying the one
quasi-particle creation operator on the even-even nuclei constructed by the BCS
theory, {\it i.e.} $| \Psi_i > = a_{i {\mu}^{'}}^+ | BCS
>$ and $| \Psi_f
> = a_{f {\nu}^{'}}^+ | BCS >$. Then the NC weak transitions  are
given by
\begin{eqnarray}
& & {\mathop\Sigma_{i {\mu}^{'} f {\nu}^{'}  }}< J_f || {\cal
{\hat O}}_{\lambda } ||J_i>  \\ \nonumber &  = & {\mathop\Sigma_{
{\mu}^{'} f {\nu}^{'}}} [ < f p || {\cal {\hat O}}_{\lambda} || i
p> ~ u_{f p {\nu}^{'}} u_{i p {\mu}^{'}} + {(-)}^{j_a + j_b +
\lambda } < i p || {\cal {\hat O}}_{\lambda} || f p> ~ v_{i p
{\mu}^{'}} v_{f p {\nu}^{'}} ] + (p \rightarrow n)~.
\end{eqnarray}
The weak current operator is comprised of longitudinal, Coulomb,
electric and magnetic operators, ${\hat O}_{\lambda }$, as described
in Ref. \cite{Ch09-2}. Finally, with the initial and final nuclear
states specified, the cross sections for $\nu ({\bar \nu}) - A$ reactions
through the weak transition operator can be directly calculated by
using the formulas of Refs. \cite{Wal75,Don79}. For CC reactions
we multiplied by the Cabbibo angle $cos ^2 \theta_c$ and took account of
the Coulomb distortion of the outgoing leptons \cite{Suzuki06,Ring08}.

\section{Results and discussions}
Here we show results for $\nu$-induced reactions for $^{92}$Nb and
$^{98}$Tc. Detailed formulae for the cross sections were presented in
our previous papers \cite{Ch09-2,Ch10}. For CC reactions, we
consider Coulomb distortion of the outgoing lepton. Since the
neutrino energies of interest here can go up to 80 MeV, we divide the
energy range into two regions. In the low energy region, we use
the Fermi correction used for the s-wave electron in $\beta$
decay. In the high energy region, however, we exploit the effective
momentum approach (EMA) used for higher energy electron scattering
analysis \cite{Co05-2,Co06}. To make a smooth transition of the cross
sections between the two energy limits, we determine an energy point, dubbed as the Coulomb cut,
below which we use the Fermi correction and above which the EMA is
used. We show results for two different Coulomb cuts, 30 and 40
MeV. Cases of 30 MeV change more smoothly. Fortunately, however,
as we show here, the temperature
dependent cross sections are nearly independent of the
Coulomb cut.

Incident $\nu ({\bar \nu})$ energies emitted  in SN explosions
\cite{Woosley90,Yoshida06} are assumed to be in the energy
range from a few to tens of MeV because the $\nu ({\bar \nu})$ energy
spectra emitted from the proto-neutron star are presumed to follow
the Fermi-Dirac distribution given by a temperature $T$ and chemical
potential $\alpha$ \cite{Yoshida06,Kolbe03-a}. Therefore, the
temperature dependence of the cross sections are averaged over the
$\nu$-distribution as follows
\begin{equation} < \sigma_{\nu} > = {\int d E_{\nu}  \sigma_{\nu} (
E_{\nu}) f ( E_{\nu}) } ~,~ f(E_{\nu}) = {  {E_{\nu}^2 } \over {
exp [( E_{\nu} / T ) - \alpha] + 1 }}~,
\end{equation}
where $\sigma_{\nu} ( E_{\nu})$ and $ f ( E_{\nu})$ are the energy
dependent $\nu - A$ cross sections and the corresponding neutrino
flux, $T$ and $\alpha$ can be chosen for a given neutrino type
\cite{Yoshida06}.

Since we are interested in  the nuclear abundances of $^{92}$Nb
and $^{98}$Tc, our temperature dependent cross sections are
presented by multiplying the particle emission branching ratios by  the
cross sections for the formation of the compound nuclei. The branching
ratios used here are based on the Hauser-Feshbach
statistical model \cite{Haus52}, using the JENDL3-3 model
of Ref. \cite{Naka05} along with the calculated transmission coefficients.
\subsection{Results for ${}^{98}$Tc}
Figure 3 shows the energy (upper) and temperature (lower panels)
dependent cross sections for CC reactions on ${}^{98}$Tc, {\it
i.e.} $^{98}$Mo$(\nu_e , e^-) ^{98}$Tc. In the upper panels, the case
of a 30 MeV Coulomb cut seems to be smoother \cite{Ring08}, so that
it seems to be more reasonable than the 40 MeV case. The final
total cross sections, however,  are nearly indistinguishable. Our energy
dependent cross sections show the typical behavior for CC cross
sections for even-even nuclei. Namely, GT $1^+$ and Fermi $0^+$
transitions dominate the total cross section below 40 MeV. However, the
contributions from  higher multipole transitions, such as
spin dipole resonance (SDR) contributions, increase above 40 MeV.

The red curves in the lower panels show the temperature dependent cross
section for the CC reaction. Different Coulomb cuts do not affect
cross sections, if we compare left and right low panels. Blue and
green curves show cross sections multiplied by the branching ratios
of the compound nuclei, $^{98}$Tc, for proton and neutron decays.
Since the neutron separation energy of ${}^{98}$Tc, $S_n$ = 7.279
MeV, is larger than the proton separation energy, $S_p$ = 6.176
MeV, proton decay is much easier than neutron decay and
leads to a larger cross section than that of the neutron decay. Of
course, these two decays have no direct relationship to the
formation of $^{98}$Tc.

Figure 4 shows the results for the energy dependent cross sections for the  NC
reactions $^{99}$Ru($\nu ({\bar \nu}), \nu^{'} ({\bar
\nu}^{'}))$$^{99}$Ru, which have no Coulomb distortion. The upper two
figures are for incident $\nu_\mu$ (left) and $\nu_e$ (right).
The lower two results are for their anti-neutrinos, {\it i.e.} ${\bar
\nu}_\mu$ (left) and ${\bar \nu}_e$ (right). One can note that
cross sections for $\nu_e$ and $\nu_\mu$ are almost identical, but those
by anti-neutrinos are different from those of incident neutrinos
if we note differences between upper and lower panels. Therefore, the NC
reactions are nearly independent of the neutrino species, but rather they depend
on the helicity of the relevant neutrinos. It is an interesting point that
the cross sections by incident $\nu$ are larger than those by incident
${\bar \nu}$ even in the case of NC reactions \cite{Ch11}. All cross sections
below 40 MeV are dominated by the GT $1^+$ transition, which is
typical of NC reactions.

Figure 5 shows the temperature dependent cross sections corresponding to
Fig.4, {\it i.e.} upper and lower two curves are for
$\nu_\mu$ (left) and $\nu_e$ (right), and ${\bar \nu}_\mu$ (left)
and ${\bar \nu}_e$ (right), respectively. Here we have only shown
results for a Coulomb cut = 40 MeV because they are nearly
independent of the cuts as shown in Fig. 3. The red curves are for the cases of
no decay, the blue and green curves include branching ratios for
neutron and proton emission decays {\it i.e.} $^{99}$Ru$( \nu (
{\bar \nu}), \nu^{'} ({\bar \nu}^{'}) ~ n)$${}^{98}$Ru and
${}^{99}$Ru$( \nu ({\bar \nu}), \nu^{'}( {\bar \nu}^{'}) ~
p)$${}^{98}$Tc, respectively. Similarly to the energy dependent cross
sections, they are independent of the neutrino species, but depend upon
neutrino helicities as can be seen in Fig. 4. These results show that,
in contrast to the results by CC, the cross sections for proton
emission decay are smaller than those for neutron emission decay
because $S_n$ = 7.464 MeV is smaller than $S_p$ = 8.478 MeV in
$^{99}$Ru.

In particular, the  green curves, $^{99}$Ru$( \nu ({\bar \nu}), \nu^{'}
({\bar \nu}^{'} ) ~p)$${}^{98}$Tc, are important contributions to
the formation of ${}^{98}$Tc. This temperature dependence could
play a vital role in understanding the neutrino temperatures at
the astrophysical sites. Since in the formation of ${}^{98}$Tc, the
nucleus may have $\beta$ decayed to ${}^{98}$Ru with a half life
of 4.2$\times 10^6$ yr and subsequently decayed to its ground
state by E2 transitions by emitting 0.74536 MeV and 0.65243 MeV
$\gamma$ rays, which might be observable just like ${}^{26}$Al in the Galaxy.
\subsection{Results for ${}^{92}$Nb}

Here we show results for the  neutrino induced reactions for
${}^{92}$Nb, whose abundance and isotopic ratio
${}^{92}$Nb/${}^{93}$Nb are of astrophysical importance because
they are thought to be produced by the $\nu$ process and can also be
used as a cosmological chronometer \cite{Yin00}.

The upper panels in Fig. 6 are the energy dependent cross sections for
$^{92}$Zr$(\nu_e , e^-) ^{92}$Nb. Likewise for the CC reaction for
${}^{98}$Tc, the 30 MeV Coulomb cut is more reasonable for the Coulomb
correction, but the location of the cut does not affect the temperature dependent cross
sections as shown in the lower panels. Since the neutron separation
energy of ${}^{92}$Nb $S_n$ = 7.883 MeV is larger than the proton
separation energy $S_p$ = 5.846 MeV, the proton decay is much
easier than the neutron decay and leads to a larger cross section
than that of neutron decay, similar to the case for
${}^{98}$Tc.

All of the results for the $\nu$-induced reactions on $^{98}$Nb
resemble  those for $^{92}$Tc. The strong energy and temperature dependence of the
cross sections, in particular the red curves in the lower panels may
give a clue to deduce the temperature conditions in the astrophysical
sites producing $^{92}$Nb. The results for $^{92}$Nb for
neutrino induced reactions via NC are presented in Fig.7, where
only results for the $\nu_e$ and ${\bar \nu}_e$ are given because they
are almost identical to those for the $\nu_\mu$ and ${\bar \nu}_\mu$
reactions.

The general trends in the energy and temperature dependent cross sections
by NC for $^{92}$Nb are shown in Fig.8. They have no special characteristics
compared to the results for $^{98}$Tc in Fig. 5 except that the cross
section for proton emission decay is larger than that for
neutron emission decay because $S_n$ = 8.831 MeV is larger than
$S_p$ = 6.043 MeV in $^{93}$Nb. The magnitudes of the cross sections are
about 1.5 times smaller than those for $^{98}$Tc. This reflects the fact that the
 cross sections for $\nu$ induced reactions are usually
proportional to the masses of the target nuclei \cite{Ch09-2,Ch10}.

\subsection{Charge exchange reactions for ${}^{98}$Tc and ${}^{92}$Nb }

We note here that the progenitor 15 $M_\odot$ model of Heger and Woosley \cite{Heger04} shows significant $^{92}$Nb presence before the supernova shock. Presumably this arises from the charge exchange reaction
${}^{92}$Mo(n,p)${}^{92}$Nb during core carbon burning. However, their network calculation does not take account of the $(n, \gamma)$ destruction of $^{92}$Nb which would happen. Therefore, we now discuss the feasibility of the formation of ${}^{98}$Tc
and ${}^{92}$Nb nuclei by the (n,p) reactions, {\it i.e.} ${}^{98}$Ru(n,p)${}^{98}$Tc
and ${}^{92}$Mo(n,p)${}^{92}$Nb, in the s-process occurring in core helium burning
during the pre-supernova evolution, and also the $(n, \gamma)$ destruction of $^{92}$Nb.

The Q value for the
former is negative $Q_{np}$ = -- 1.014 MeV, so that even neutrons with energies around a few hundred keV cannot be captured to produce
${}^{98}$Tc. However,  the ${}^{92}$Mo(n,p)${}^{92}$Nb reaction, whose
Q value is $Q_{np}$ = 0.42671 MeV, may  occur for neutrons at $E
\sim$ 30 -- 100 keV, if we consider the following discussion.

The $J^{\pi}$ for states below $\sim$ 0.5 MeV in  $^{92}$Nb are $J^{\pi} = 7^+(0.0) $,  $2^+
(0.135),$  $2^- (0.225),$  $3^+ (0.285),$ $ 5^+ (0.353),$ $ 3^- (0.390),$ $4^+ (0.480)$ and $
6^+ (0.501)$. Therefore, two lowest states of the $^{92}$Nb + p system are
$J^{\pi} = {15 \over 2}^+$ or ${13 \over 2}^{+}$ for the ground and $J^{\pi} =
{5 \over 2}^+$ or ${3 \over 2}^{+}$ for the 1st excited states.
Since the initial system, $^{92}$Mo + n, is given as $0^+ \otimes
{ 1\over 2}^+ \otimes l_{n}^{\pi}$, we need at least p-wave
neutrons, ${\it i.e.}$ $l_n^{\pi} \ge 1^-$.

Even if we consider the excited states of $^{92}$Nb below $\sim$ 0.5 MeV,
which could be populated at a stellar temperature of $T_9 = T/10^9 K \approx 0.3$
for the s-process ({\it i.e.} typical neutron energy  $\sim$ 30 keV)
because $E_n(=30 keV) + Q_{np}(=0.43 MeV) \simeq $ 0.5 MeV,
only the $2^-$ (0.225) and $3^-$ (0.389) states are allowed with $l_p^{\pi} =
1^-$ and $2^+$ protons by the conservation of relevant angular momenta.

We do not have any experimental data for the reaction ${}^{92}$Mo(n,p)${}^{92}$Nb
below 1.5 MeV. According to theoretical calculations
by ENDF/B-VII.0 \cite{Chadwick}, however, this reaction cross section
might be lower than 0.1 $\mu b$ for neutrons at energies below 1 MeV.
Therefore the cross sections at about 30 keV neutrons might be
much smaller than 0.1 $\mu b$.
We carried out a HF statistical model calculation
at the threshold energy region
as employing the JENDL-4 data \cite{Shibata}.
The calculated Maxwellian-averaged cross sections for ${}^{92}$Mo(n,p)${}^{92}$Nb
turn out to be extremely small $ {<\sigma v>} / {v_T} = 4.02 \times 10^{-13} \mu b$
and $5.39 \times 10^{-4} \mu b$
at the neutron energies 30 keV and 100 keV, respectively,
where $< ~ >$ denotes an average with respect to the Maxwellian spectrum,
$\sigma$ is the cross section, $v$ is the relative velocity of the neutrons and target nucleus,
and $v_T$ is the mean thermal velocity.

Once ${}^{92}$Nb is produced by the (n,p) reaction from ${}^{92}$Mo,
it is exposed simultaneously to an intense flux of neutrons and destroyed by
the radiative neutron capture reaction ${}^{92}$Nb(n,$\gamma$)${}^{93}$Nb.
Although the (n,$\gamma$) cross section was not measured
for the radioactive nucleus ${}^{92}$Nb ($\tau_{1/2}=3.47 \times10^7 y$),
the ${}^{92}$Nb(n,$\gamma$)${}^{93}$Nb cross section is expected to be as large as
those measured for stable Nb isotopes,
$ {<\sigma v>} / {v_T} = 261.3,~317.2$, and $402.6~mb$ for
${}^{93,94,95}$Nb(n,$\gamma$)${}^{94,95,96}$Nb reactions, respectively, at the neutron energy 30 keV \cite{Naka05}.
These (n,$\gamma$) cross sections are eighteen orders of magnitude larger than the
${}^{92}$Mo(n,p)${}^{92}$Nb cross section at this energy.
Therefore, the ${}^{92}$Mo(n,p)${}^{92}$Nb reaction should
not contribute much to the production of $^{92}$Nb
in the weak s-process in core helium burning phase of massive stars
before explosion.

\section{Summary}
We have calculated neutrino induced reactions on two odd-odd nuclei,
$^{98}$Tc and $^{92}$Nb by the Quasi-particle RPA method because both
nuclei may be produced by the $\nu$ process in the explosive
astrophysical objects. The abundance of $^{98}$Tc can be measured by
observing the $\gamma$-ray from the E2 transition to $^{98}$Ru.
Consequently, it may play a role as another $\gamma$-ray source
for astrophysical observation similar to $^{26}$Al.

The abundance ratio of $^{92}$Nb/$^{93}$Nb has a meaningful implication.
The ratio could determine the various roles of neutrino
properties in explosive nucleosynthesis because it is very
rare compared to that of $^{138}$La and $^{180}$Ta whose isotopic
abundances place valuable physical constraints on the
production site \cite{Hayakawa10}.

The energy and temperature dependent cross sections for ${}^{92}$Nb,
$^{92}$Zr($\nu_e,e^{-}$)$^{92}$Nb by charged current (CC) and
$^{93}$Nb(${\nu},~{\nu}^{'}$ n )$^{92}$Nb by neutral current (NC)
are presented. For ${}^{98}$Tc, the CC reaction
$^{98}$Mo($\nu_e,e^-$)$^{98}$Tc and the NC reaction
$^{99}$Ru(${\nu},~{\nu}^{'}$ p)$^{98}$Tc have been estimated using the
QRPA. Particle emission decays of the compound nuclei produced by
the $\nu$ process make use of branching ratios
calculated in the framework of the Hauser-Feshbach statistical approach
with theoretical transmission coefficients.

Our deduced  cross sections for these nuclei show features
typical of neutrino induced reactions by NC and CC. CC reactions
are dominated by GT transition below 40 MeV, but other
multipole transitions become large above that energy. In
the case of the NC reactions, the GT dominance becomes more significant in
the low energy region. One more point of note regarding the NC reactions
is that neutrino reactions are nearly independent of
neutrino species, but depend on the neutrino helicity.

Finally,  discussions of the charge exchange reaction by thermal neutrons
for these nuclei remind us that the $^{92}$Mo(n,p)$^{92}$Nb
reaction might affect the initial abundance ratio of
$^{92}$Nb/$^{93}$Nb before the explosion, while $^{98}$Ru(n,p)$^{98}$Tc
may not because of the negative $Q_{np}$ value.
However, the (n,p) reaction is expected to be impotent for the production
of pre-existing $^{92}$Nb because the ${}^{92}$Nb(n,$\gamma$)${}^{93}$Nb
reaction cross section is $\sim 10^{18}$ times larger than that of $^{92}$Nb/$^{93}$Nb
for neutrons of the energy $\sim$ 30 keV and such $^{92}$Nb is quickly destroyed.
Nevertheless,  more thorough calculations are necessary
before  further
decisive conclusions about the roles of charge exchange reactions for
$^{92}$Nb can be made. Nucleosyntheses in SN explosions which consider
these neutrino reactions are in progress including  more realistic
calculations of the charge exchange reactions.

\acknowledgments

This work was supported by the National Research Foundation of
Korea (2011-0003188) and one of authors, Cheoun, was supported by
the Soongsil University Research Fund. This work was also supported in part by
Grants-in-Aid for Scientific Research of JSPS (20244035),
and in part by U.S. Department of Energy under Nuclear Theory Grant DE-FG02-95-ER40934,
and in part by Grants-in-Aid for JSPS Fellows (21.6817)

\newpage
\begin{figure}
\centering
\includegraphics[width=12.5cm]{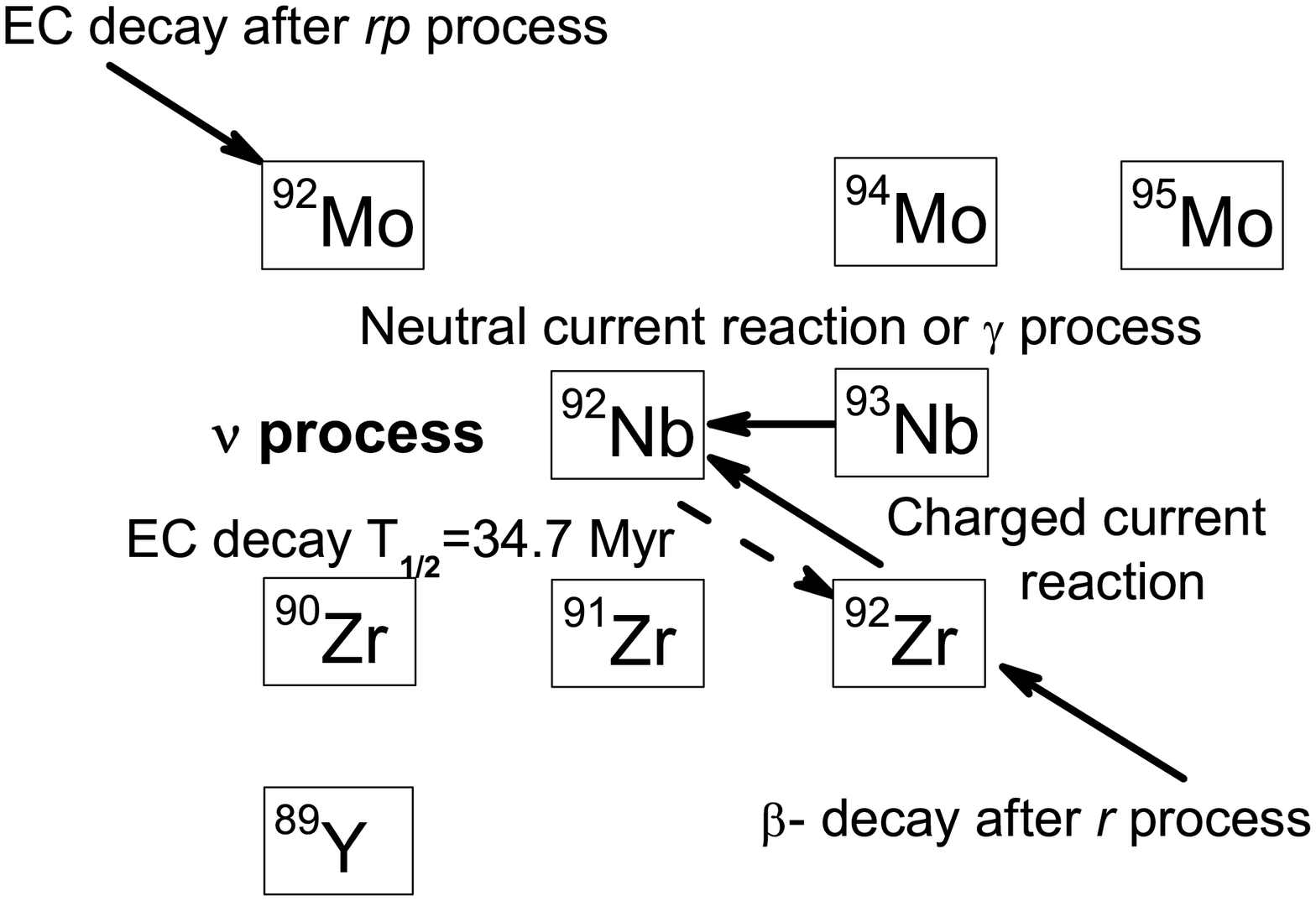}
\caption{Partial nuclear chart around $^{92}$Nb and the main neutrino
reactions, $^{92}$Zr($\nu_e,e^{-}$)$^{92}$Nb by charged
current(CC) and $^{93}$Nb(${\nu} ({\bar \nu}),~{\nu}^{'} ({\bar
\nu})^{'} $ n )$^{92}$Nb by neutral current(NC).} \label{fig.l}
\end{figure}

\newpage
\begin{figure}
\centering
\includegraphics[width=12.5cm]{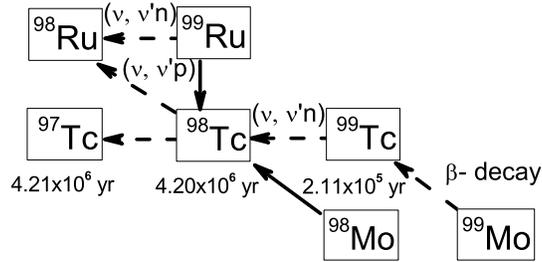}
\caption{Partial nuclear chart around $^{98}$Tc and main neutrino
reactions, $^{98}$Mo($\nu_e,e^-$)$^{98}$Tc by CC and
$^{99}$Ru(${\nu} ({\bar \nu}) ,~{\nu}^{'} ({\bar \nu})^{'}$
p)$^{98}$Tc by NC.} \label{fig.2}
\end{figure}

\newpage
\begin{figure}
\centering
\includegraphics[width=7.5cm]{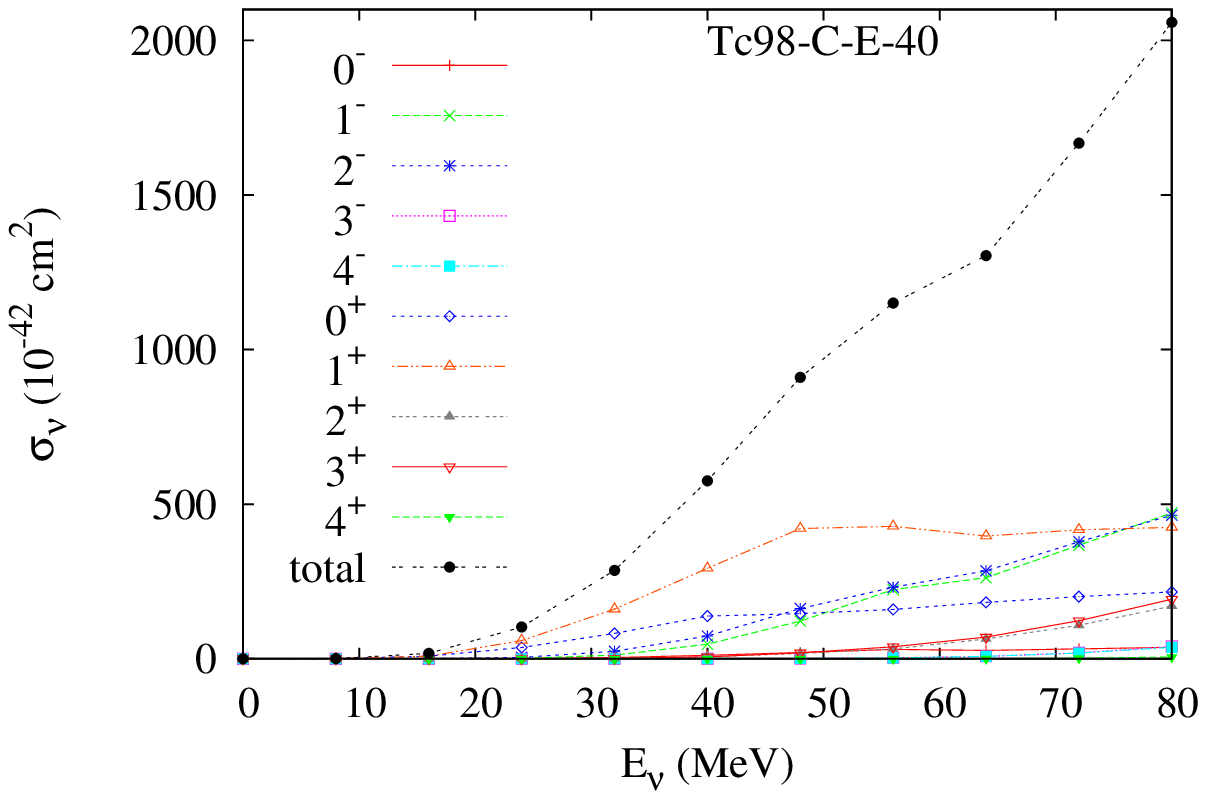}
\includegraphics[width=7.5cm]{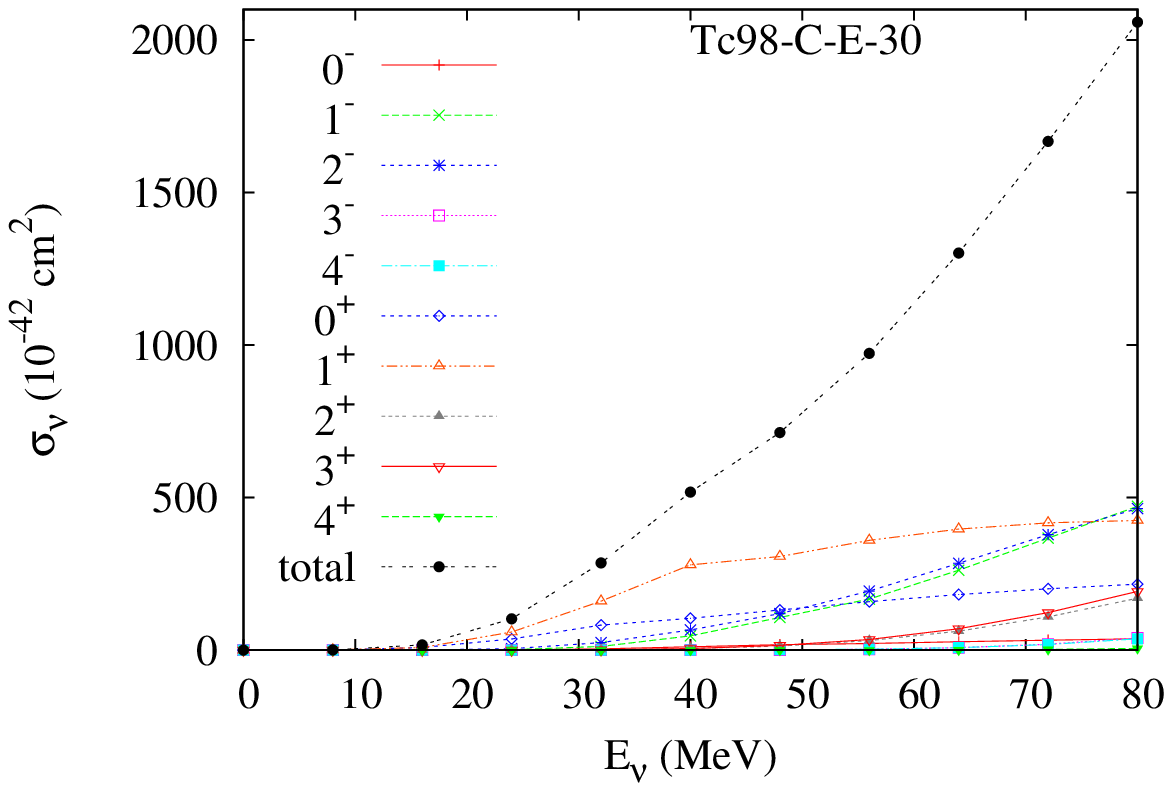}
\includegraphics[width=7.5cm]{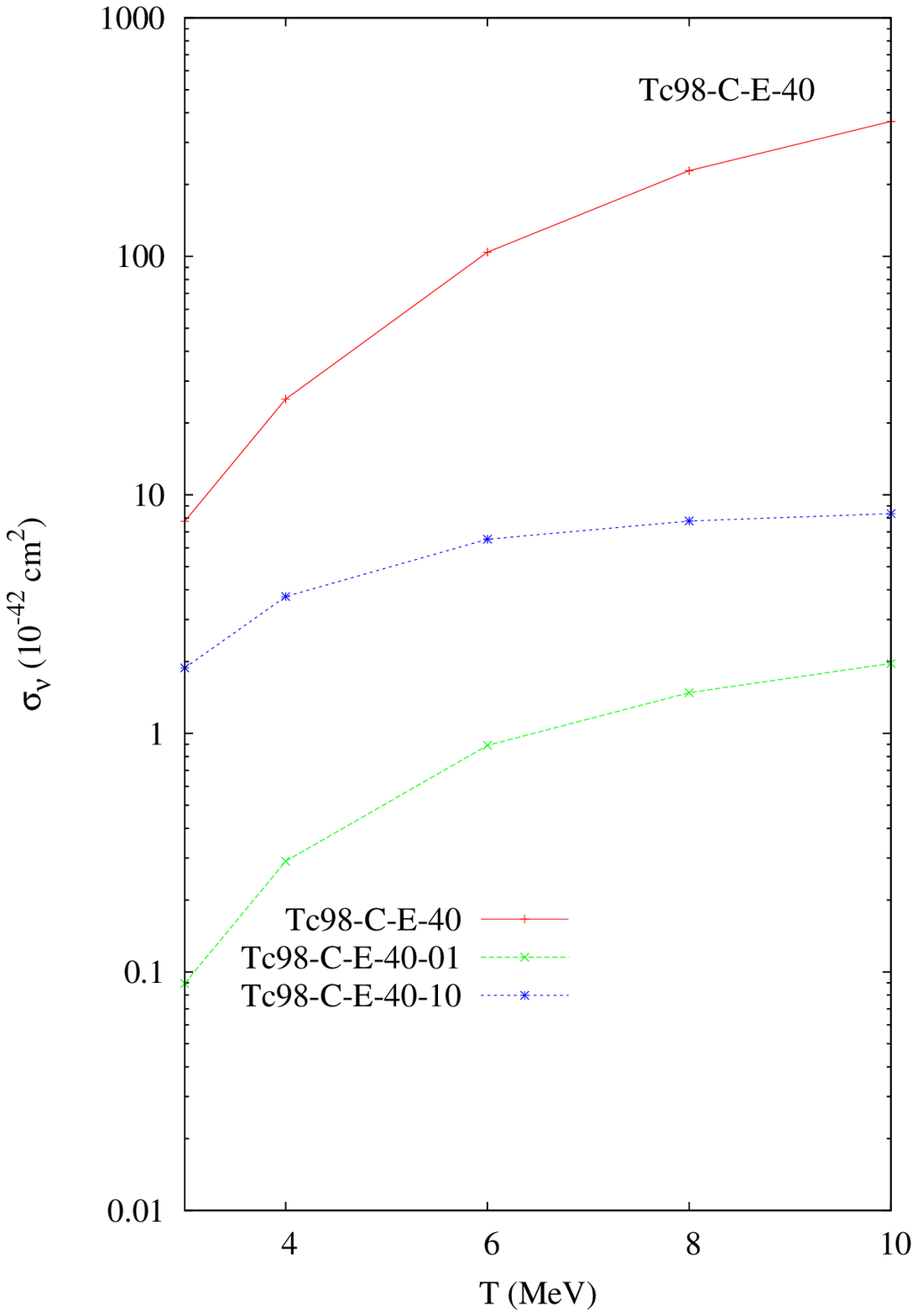}
\includegraphics[width=7.5cm]{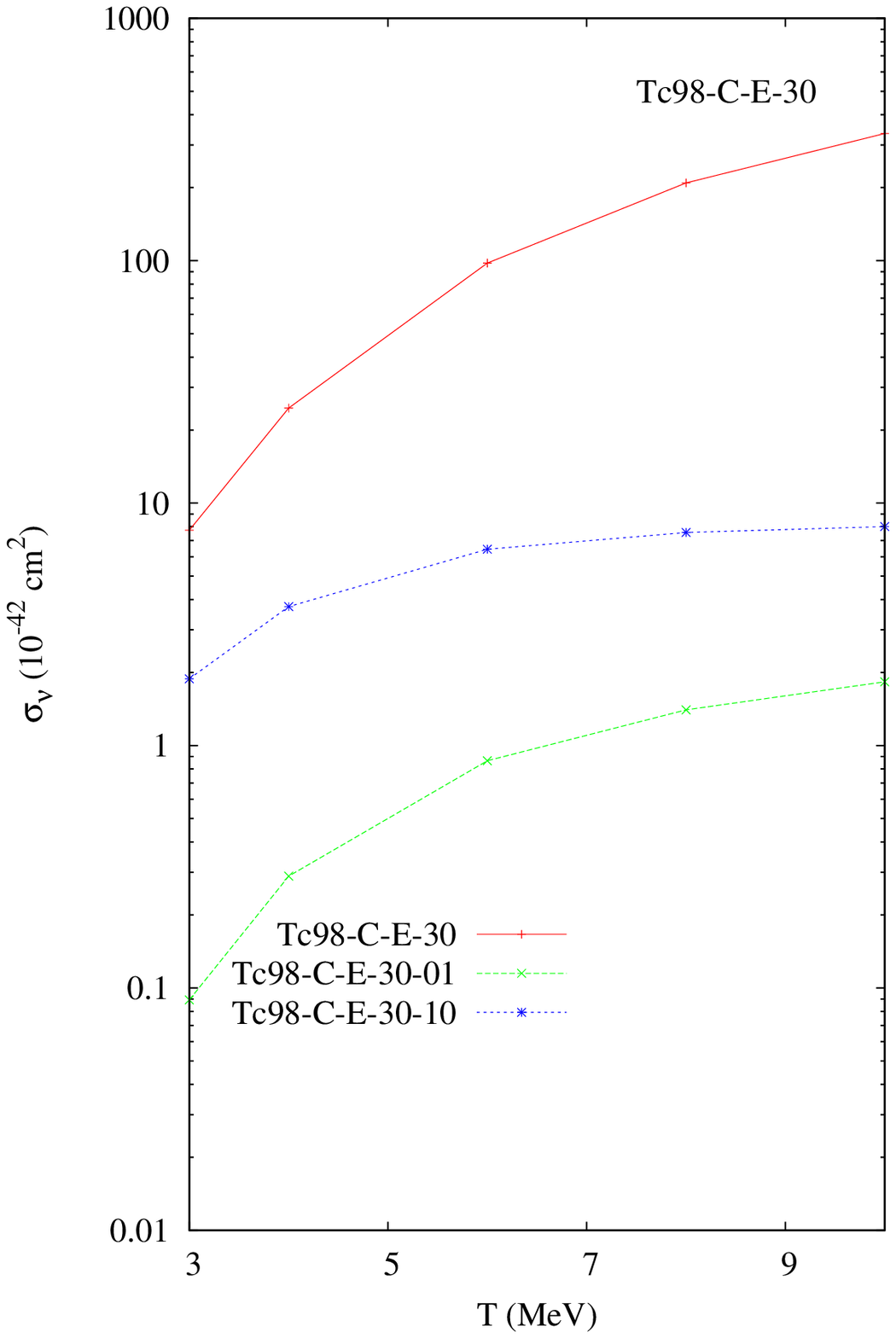}
\caption{Energy (upper) and temperature (lower) dependent cross
sections of CC reaction for ${}^{98}$Tc, ${}^{98}$Mo$( \nu_e, e^-
) ^{98}$Tc. The contribution of each multipole transition is also
presented along with their sum. Left and right panels are for Coulomb
cuts = 40 and 30 MeV (see text for more explanations),
respectively. The red curves in the lower panels are cross sections for
${}^{98}$Tc, ${}^{98}$Mo$( \nu_e, e^- ) ^{98}$Tc. Blue and green
curves are for proton and neutron emission decays from
${}^{98}$Tc$^*$, {\it i.e.} ${}^{98}$Mo$( \nu_e, e^- p)$${}^{97}$Mo and
$^{98}$Mo$( \nu_e, e^- n)$${}^{97}$Tc.} \label{fig.3}
\end{figure}

\newpage
\begin{figure}
\centering
\includegraphics[width=7.5cm]{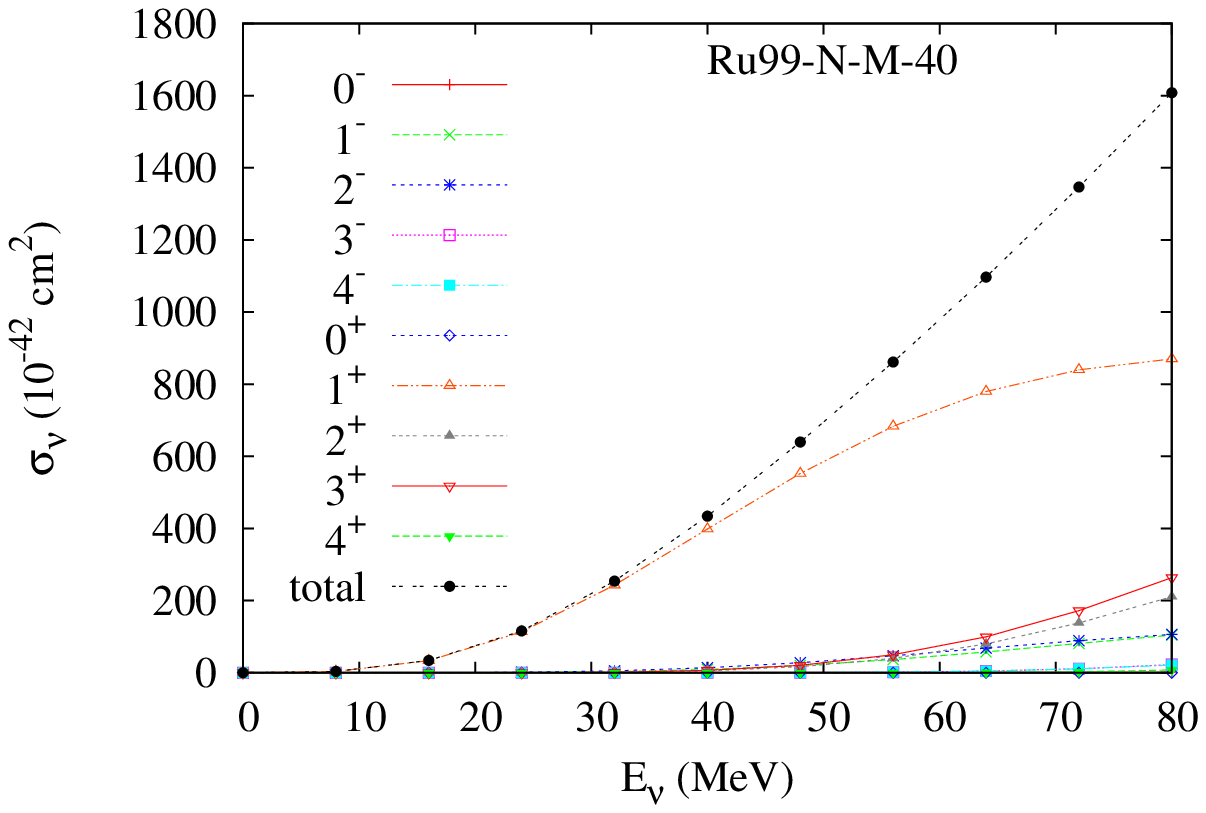}
\includegraphics[width=7.5cm]{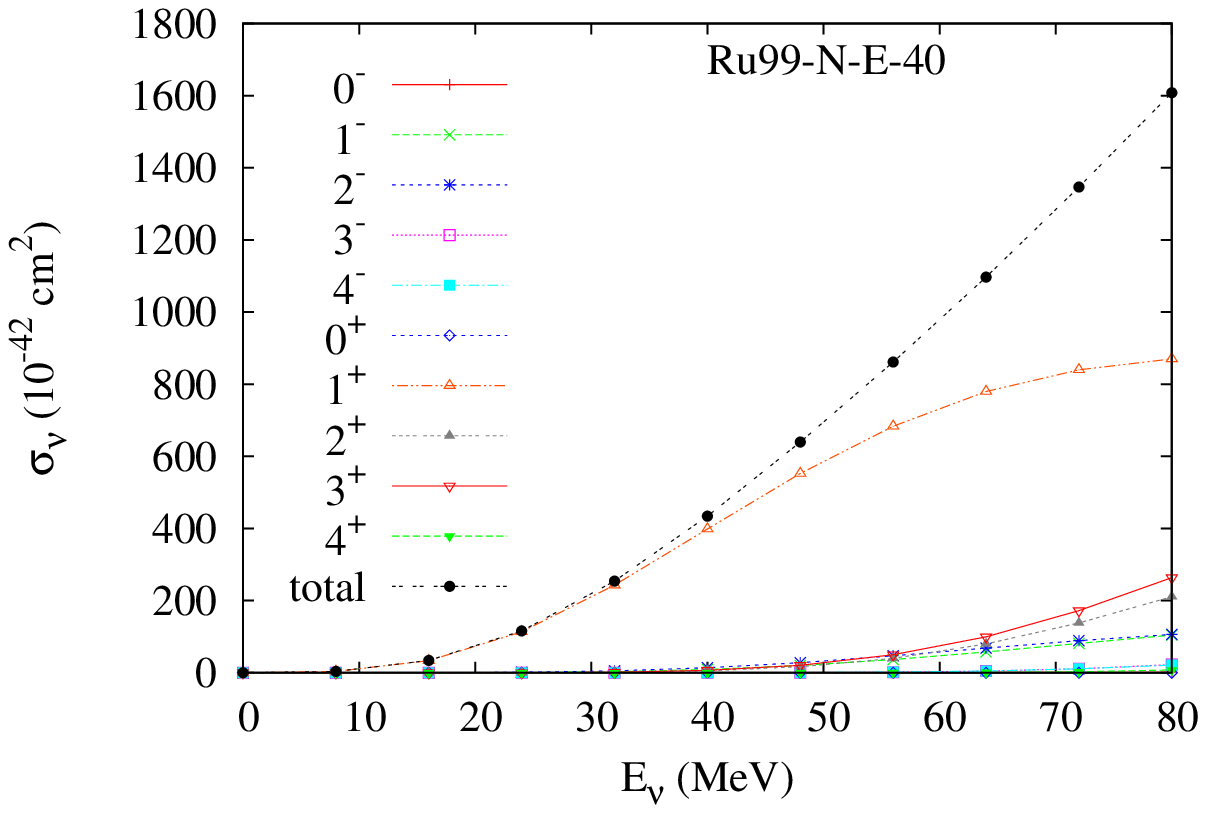}  \\
\includegraphics[width=7.5cm]{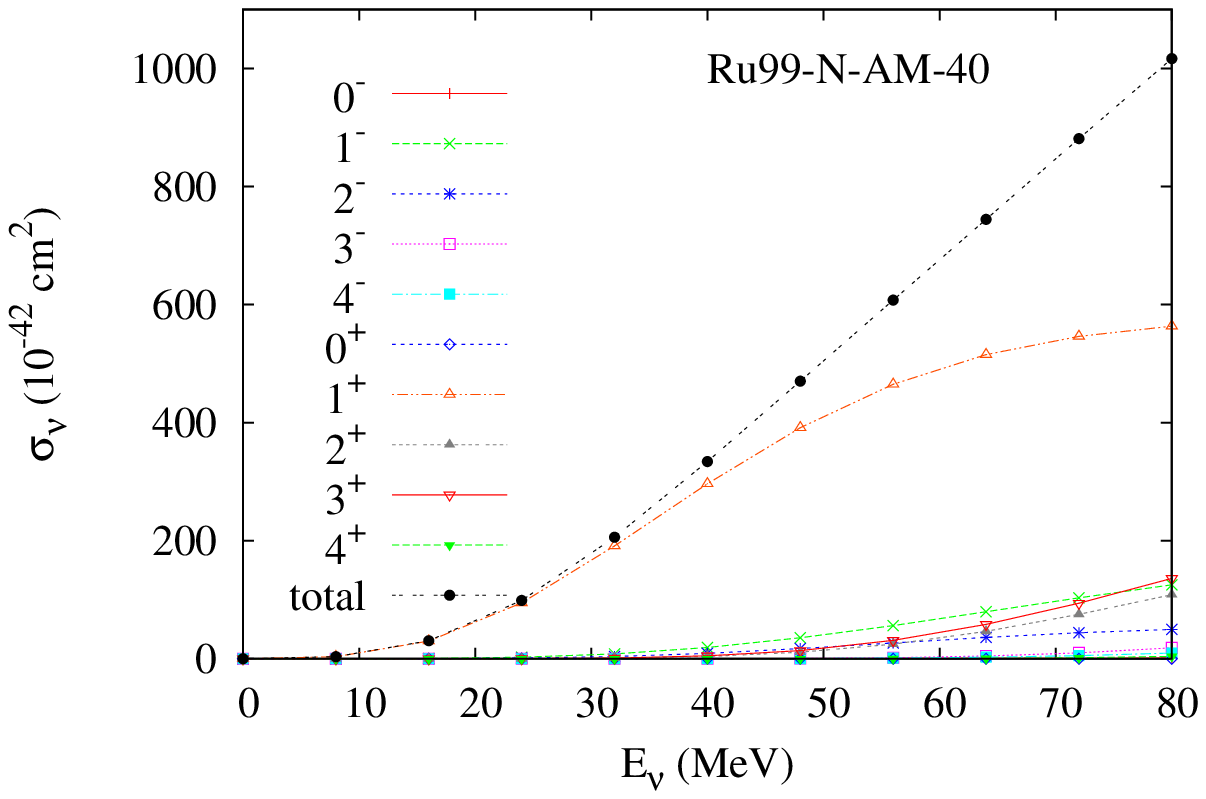}
\includegraphics[width=7.5cm]{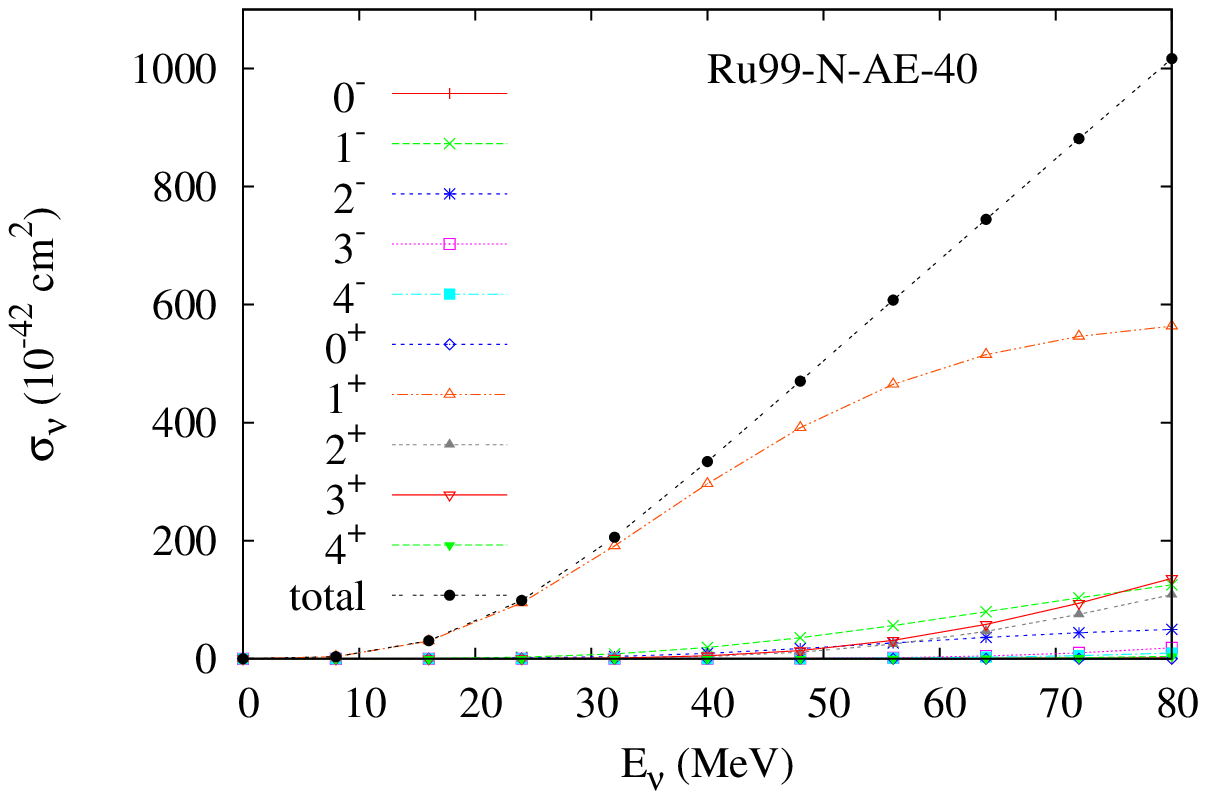}
\caption{Energy dependent cross sections of NC reactions for
${}^{98}$Tc, $^{99}$Ru(${\nu} ({\bar \nu}) ,~{\nu}^{'} ({\bar
\nu})^{'}$ )$^{99}$Ru. The upper two figures are for incident
$\nu_\mu$ (left) and $\nu_e$ (right). The lower two results are for
${\bar \nu}_\mu$ (left) and ${\bar \nu}_e$ (right).} \label{fig.4}
\end{figure}

\newpage
\begin{figure}
\centering
\includegraphics[width=7.2cm]{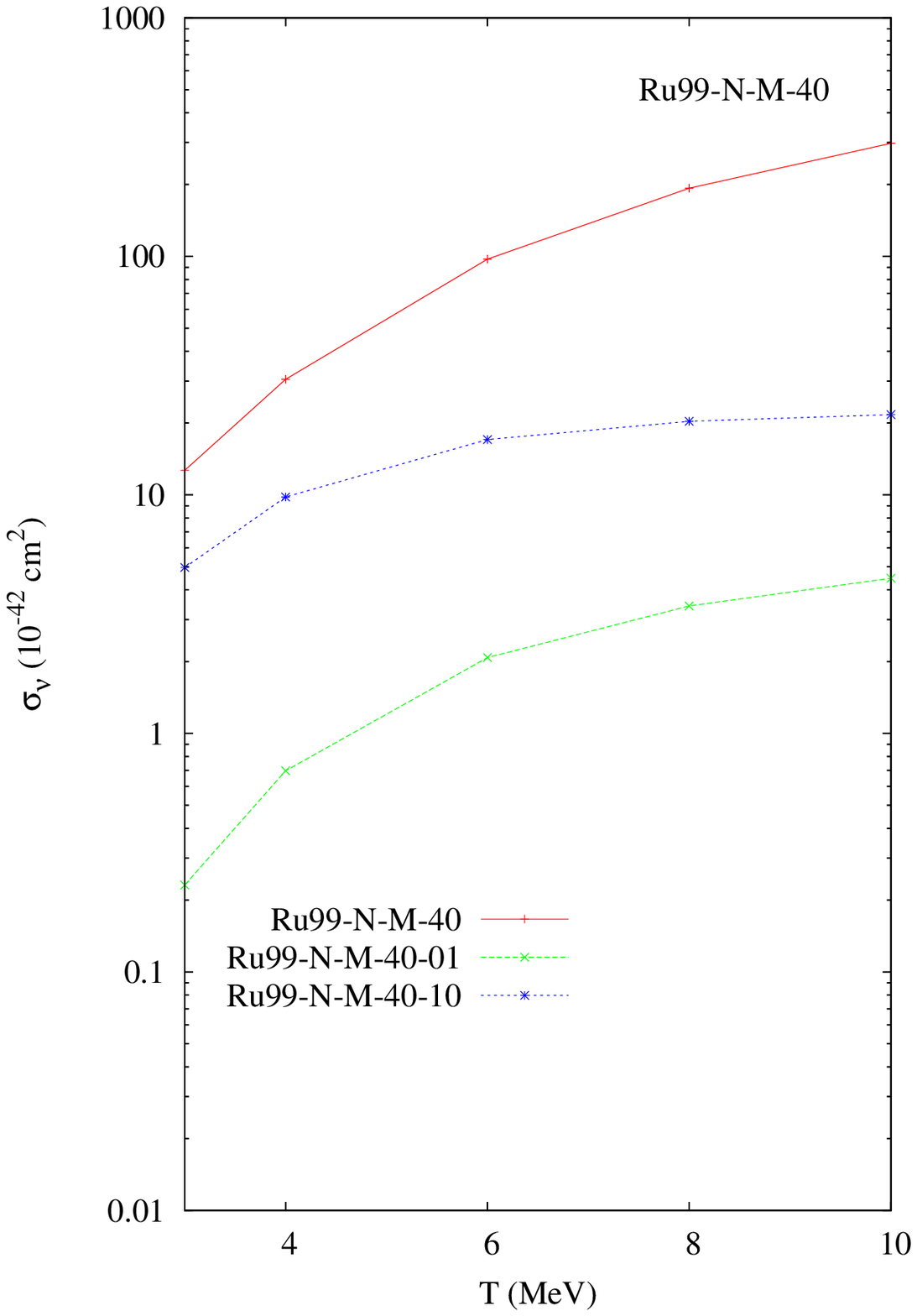}
\includegraphics[width=7.2cm]{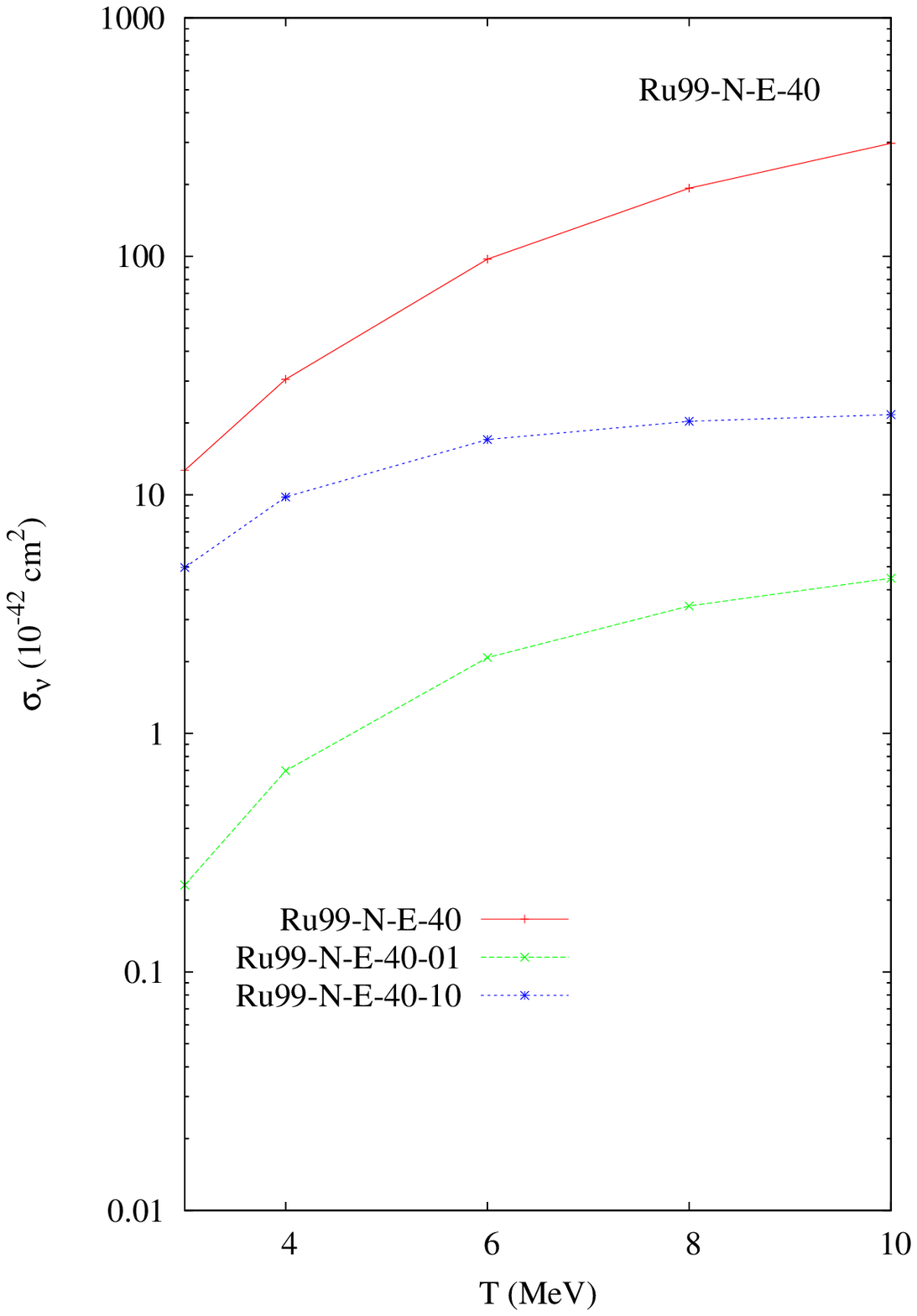}  \\
\includegraphics[width=7.2cm]{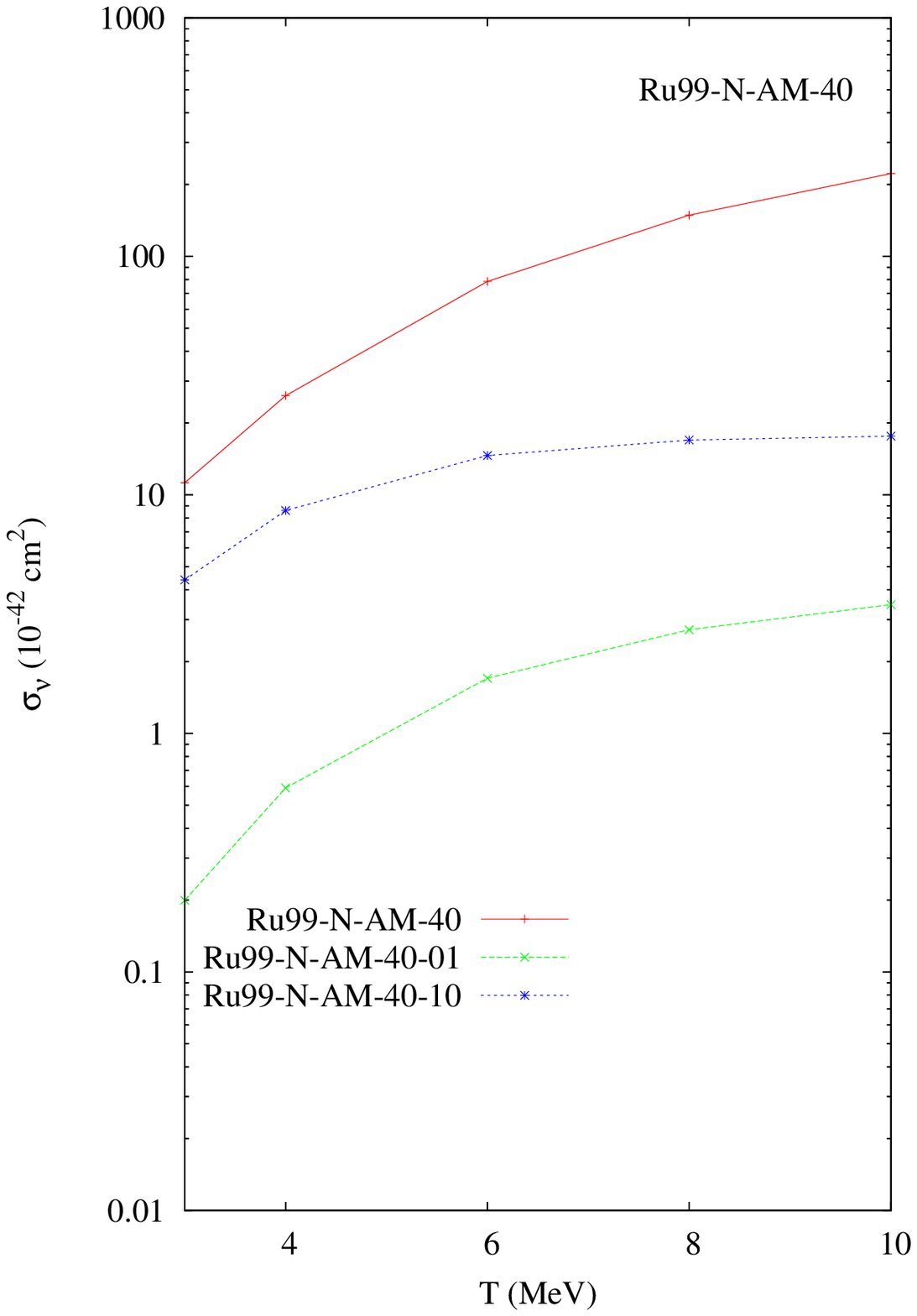}
\includegraphics[width=7.2cm]{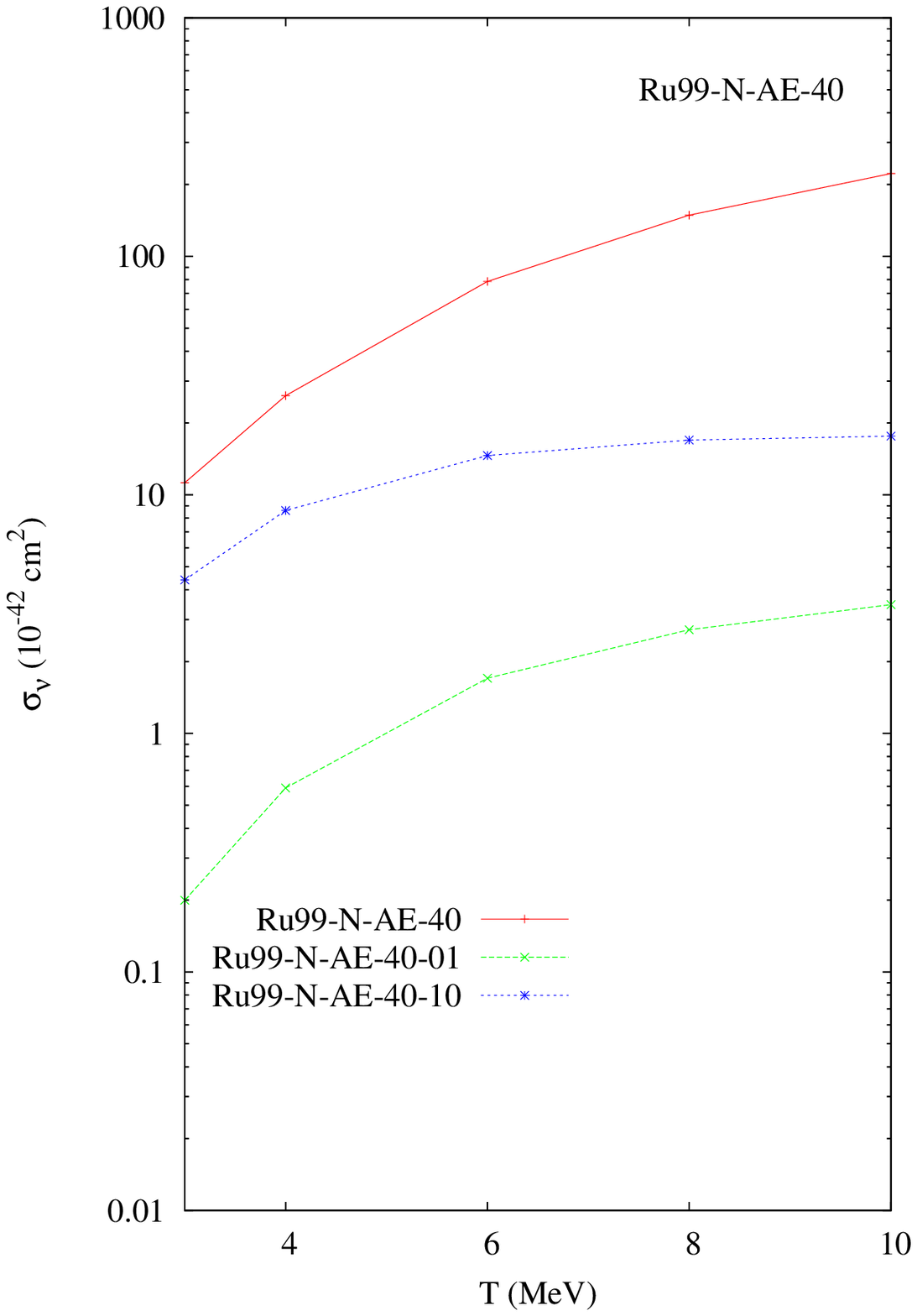}
\caption{Temperature dependent cross sections of NC reactions for
${}^{98}$Tc. The upper two figures are for incident $\nu_\mu$ (left)
and $\nu_e$ (right). The lower two results are for ${\bar \nu}_\mu$
(left) and ${\bar \nu}_e$ (right). Red curves are ${}^{99}$Ru$(
\nu ({\bar \nu}), \nu^{'} ({\bar \nu^{'}}) )$${}^{99}$Ru. Blue and
green curves are for neutron and proton emission decays from
${}^{99}$Tc$^*$, {\it i.e.} $^{99}$Ru$( \nu ({\bar \nu}), \nu^{'}
({\bar \nu^{'}}) n)$${}^{98}$Ru and ${}^{99}$Ru$(
 \nu ({\bar \nu}),
\nu^{'} ({\bar \nu^{'}}) p)$${}^{98}$Tc.} \label{fig.5}
\end{figure}


\newpage
\begin{figure}
\centering
\includegraphics[width=7.5cm]{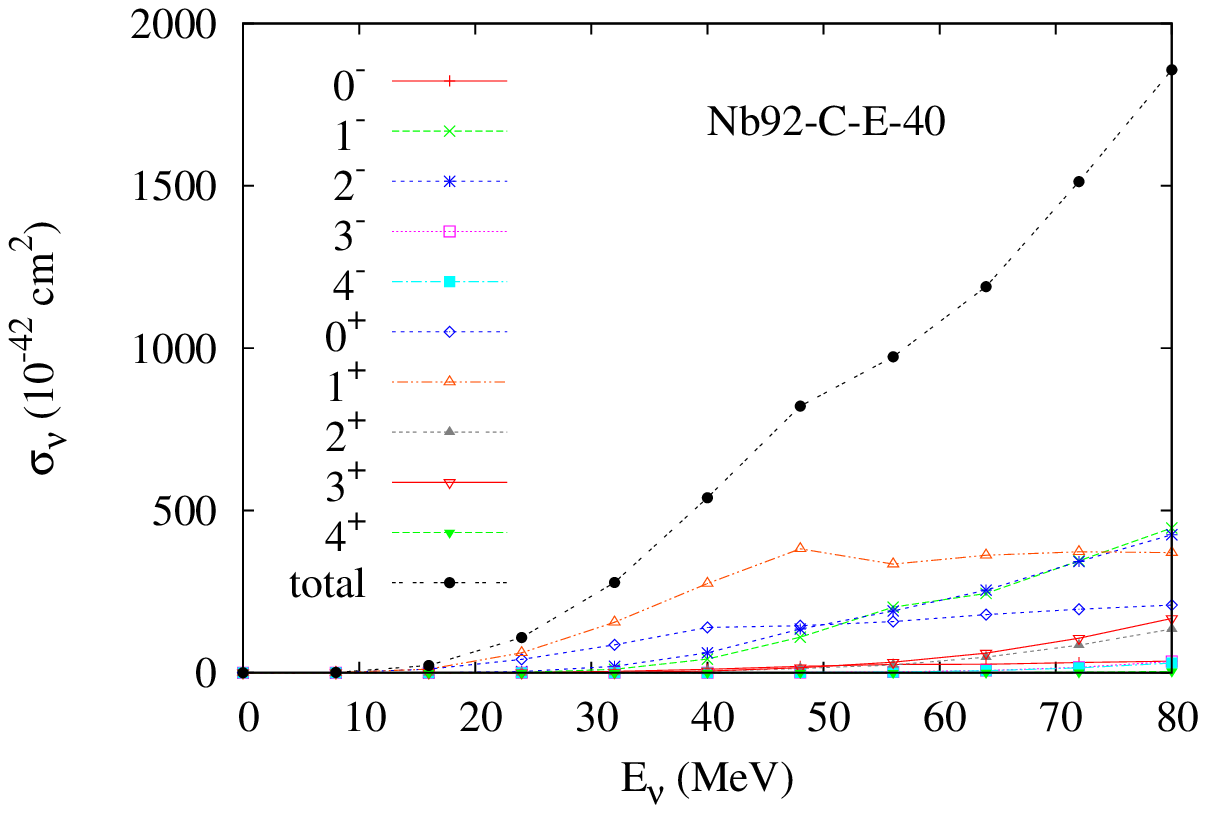}
\includegraphics[width=7.5cm]{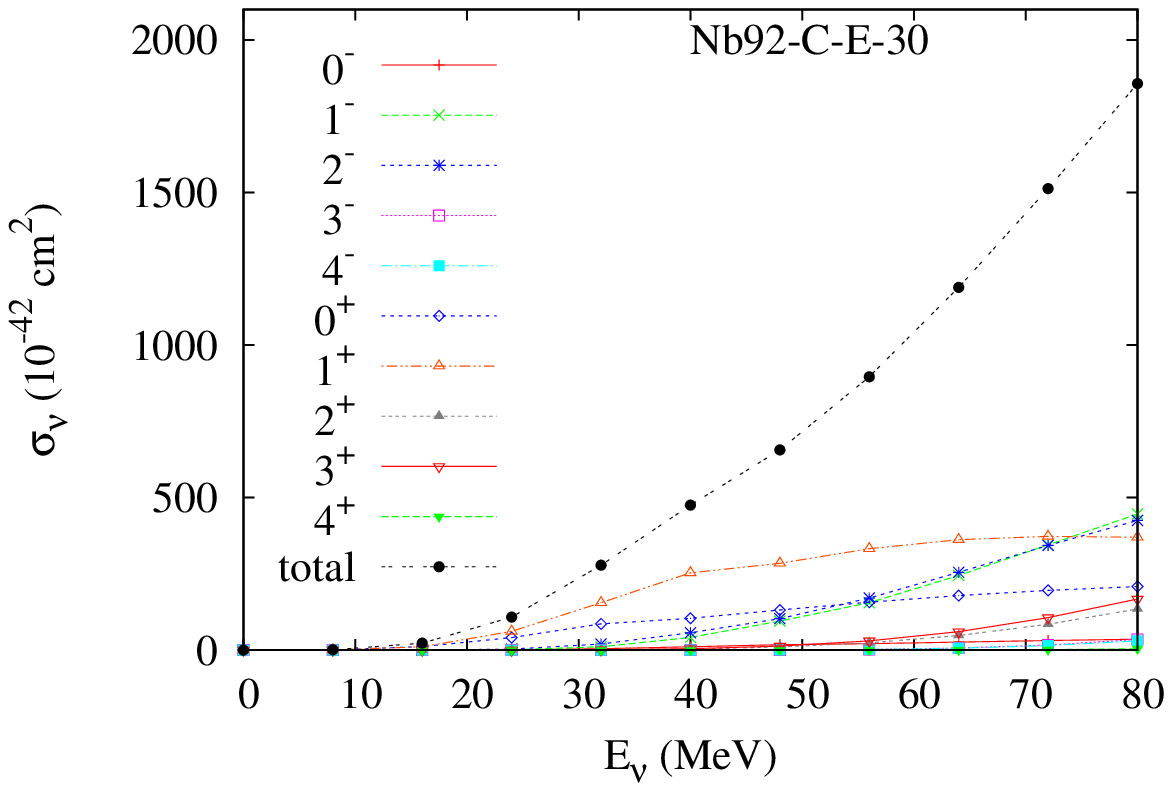}  \\
\includegraphics[width=7.5cm]{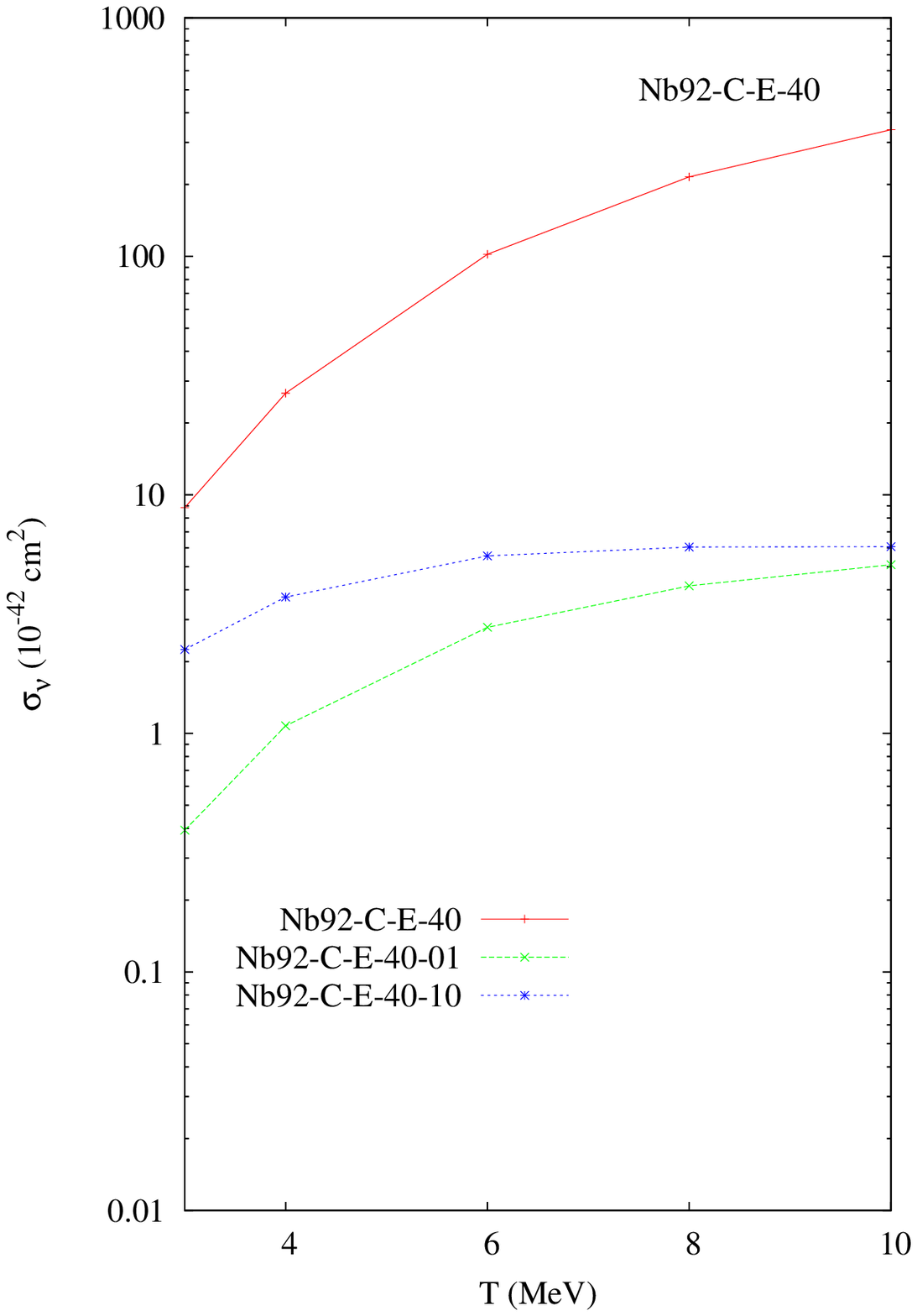}
\includegraphics[width=7.5cm]{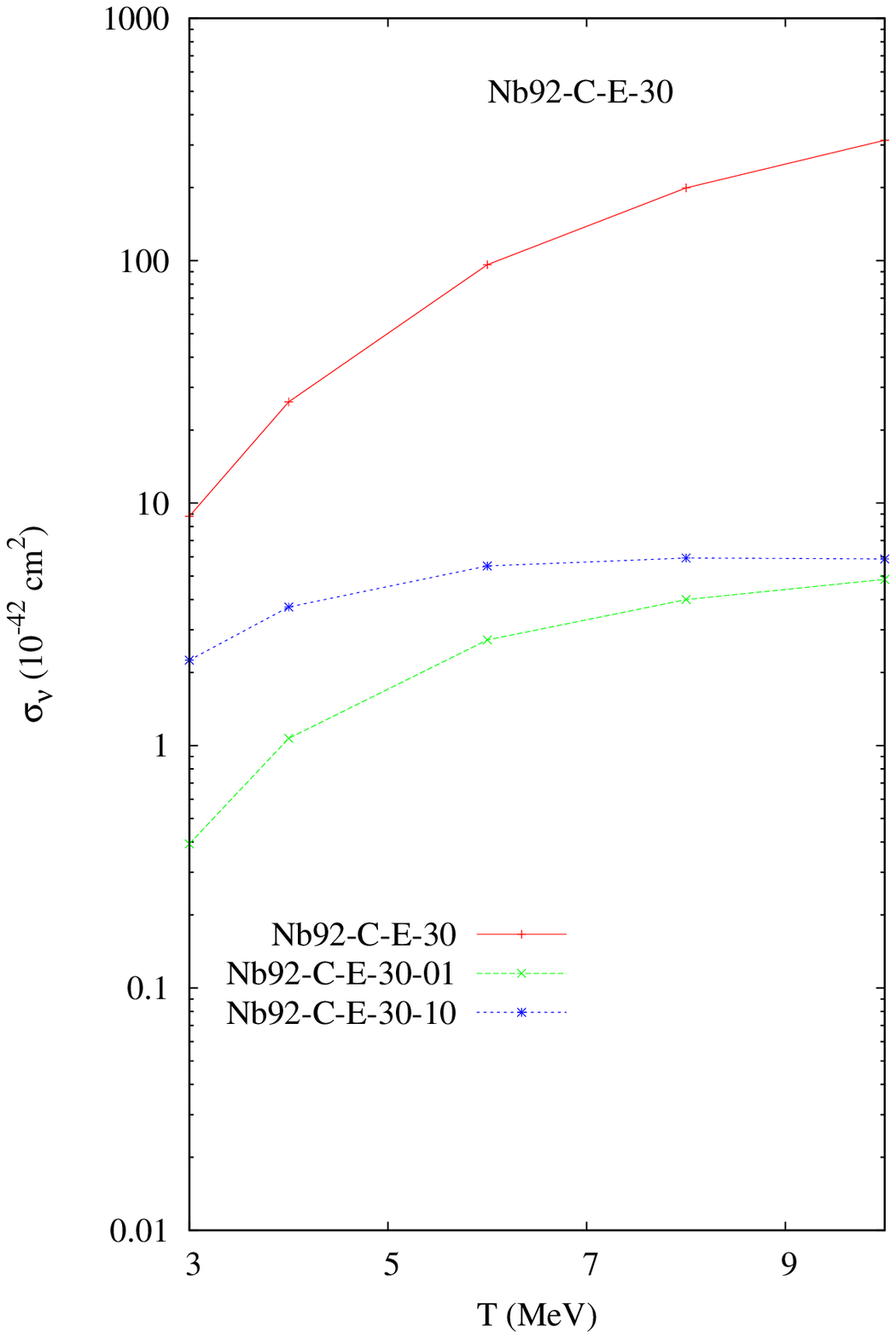}
\caption{Energy and temperature dependent cross sections of CC
reactions for ${}^{92}$Nb. Red curves in lower panels are cross
sections for ${}^{92}$Nb, ${}^{92}$Zr$( \nu_e, e^- ) ^{92}$Nb.
Blue and green curves are for proton and neutron decays, {\it i.e.}
${}^{92}$Zr$( \nu_e, e^- p)$${}^{91}$Zr and $^{92}$Zr$( \nu_e, e^-
n)$${}^{91}$Nb.} \label{fig.6}
\end{figure}
\newpage
\begin{figure}
\centering
\includegraphics[width=7.5cm]{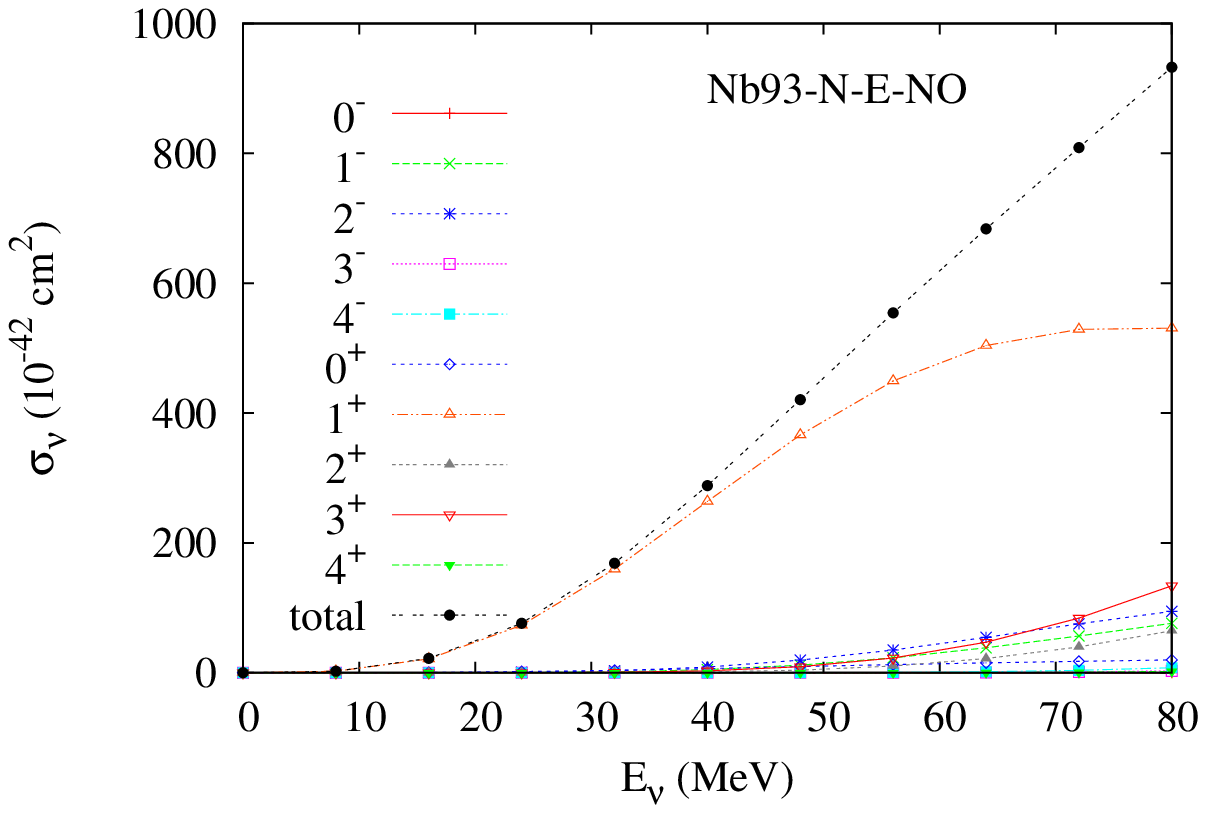}
\includegraphics[width=7.5cm]{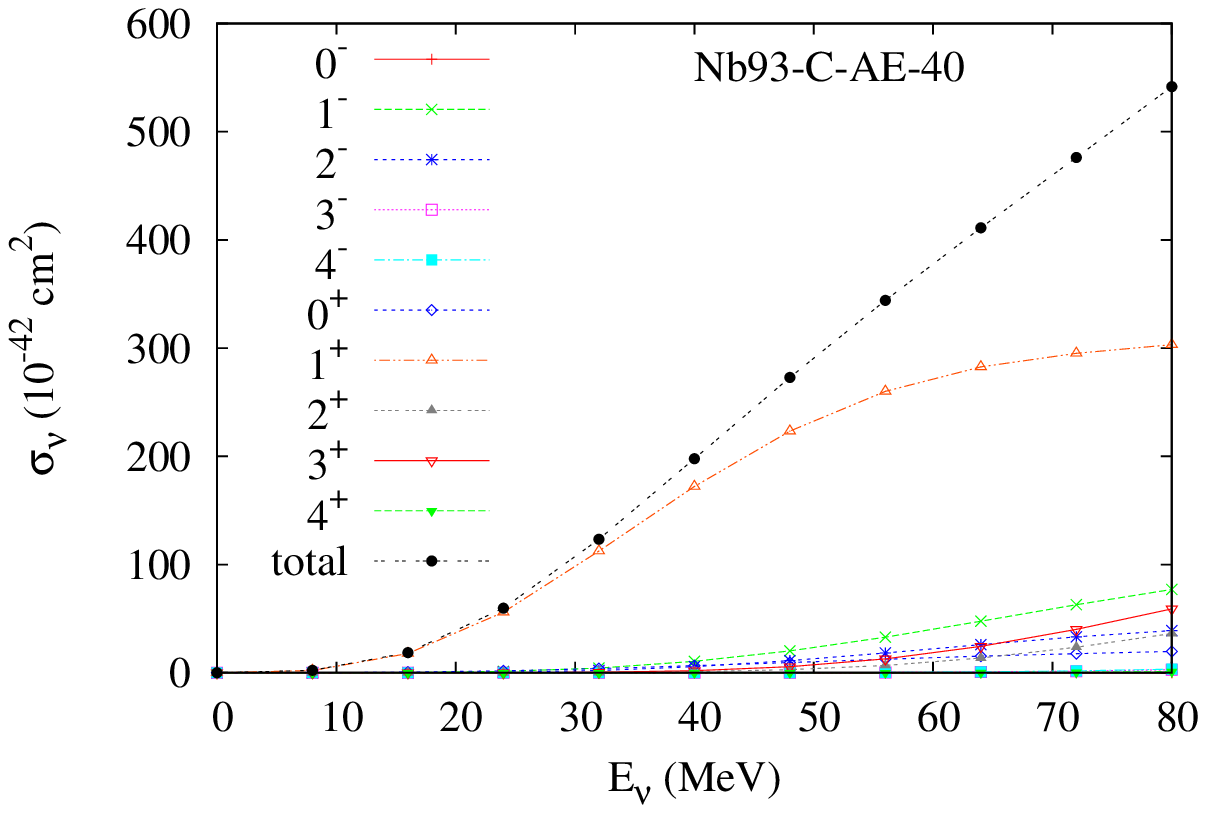}  \\
\caption{Energy dependent cross sections of NC reactions for
${}^{92}$Nb, $^{93}$Nb$( \nu ({\bar \nu}), \nu^{'} ({\bar
\nu^{'}}) n)$${}^{93}$Nb. Left is for incident $\nu_e$ and right
is for ${\bar \nu}_e$.} \label{fig.7}
\end{figure}

\begin{figure}
\centering
\includegraphics[width=7.2cm]{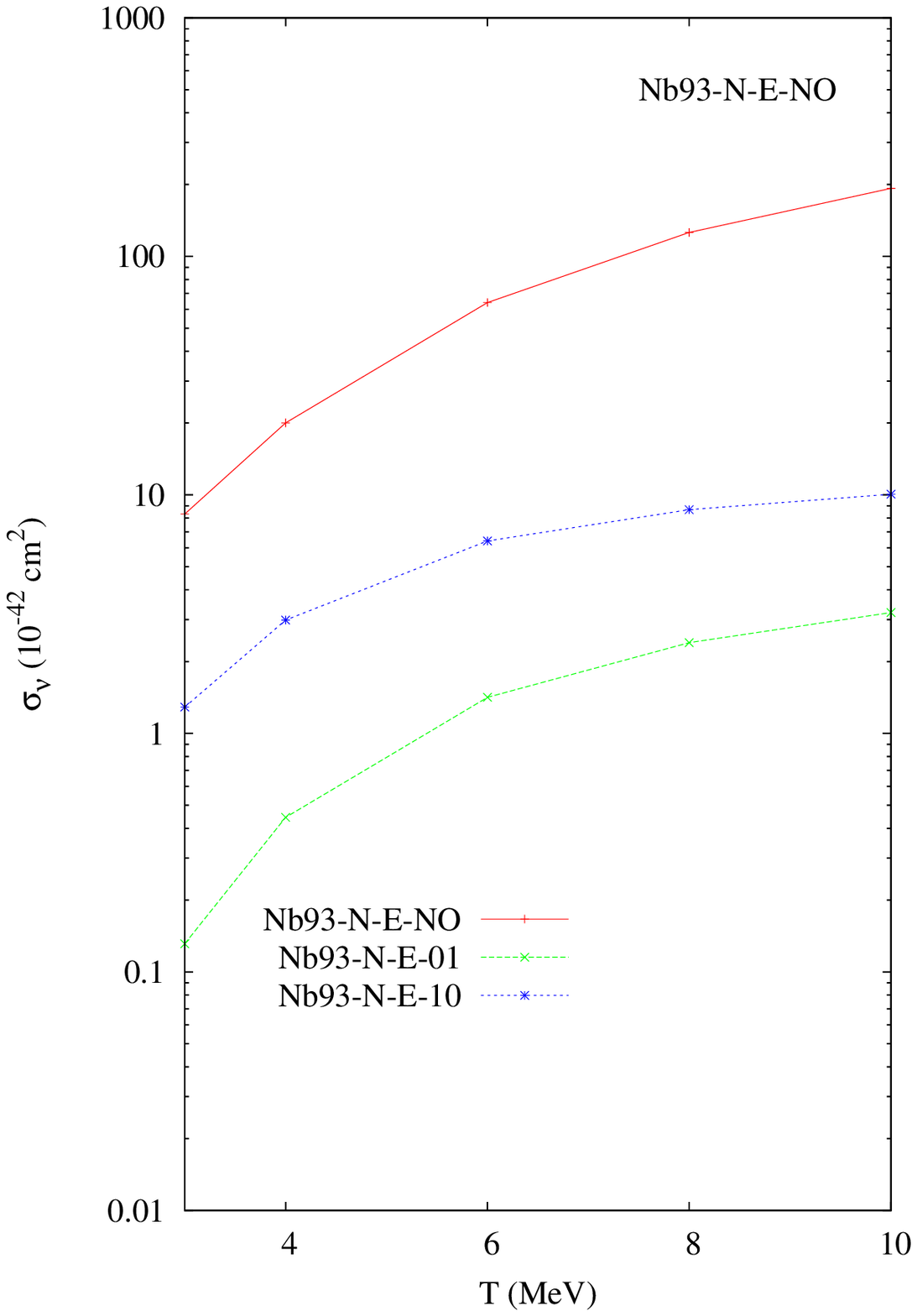}
\includegraphics[width=7.2cm]{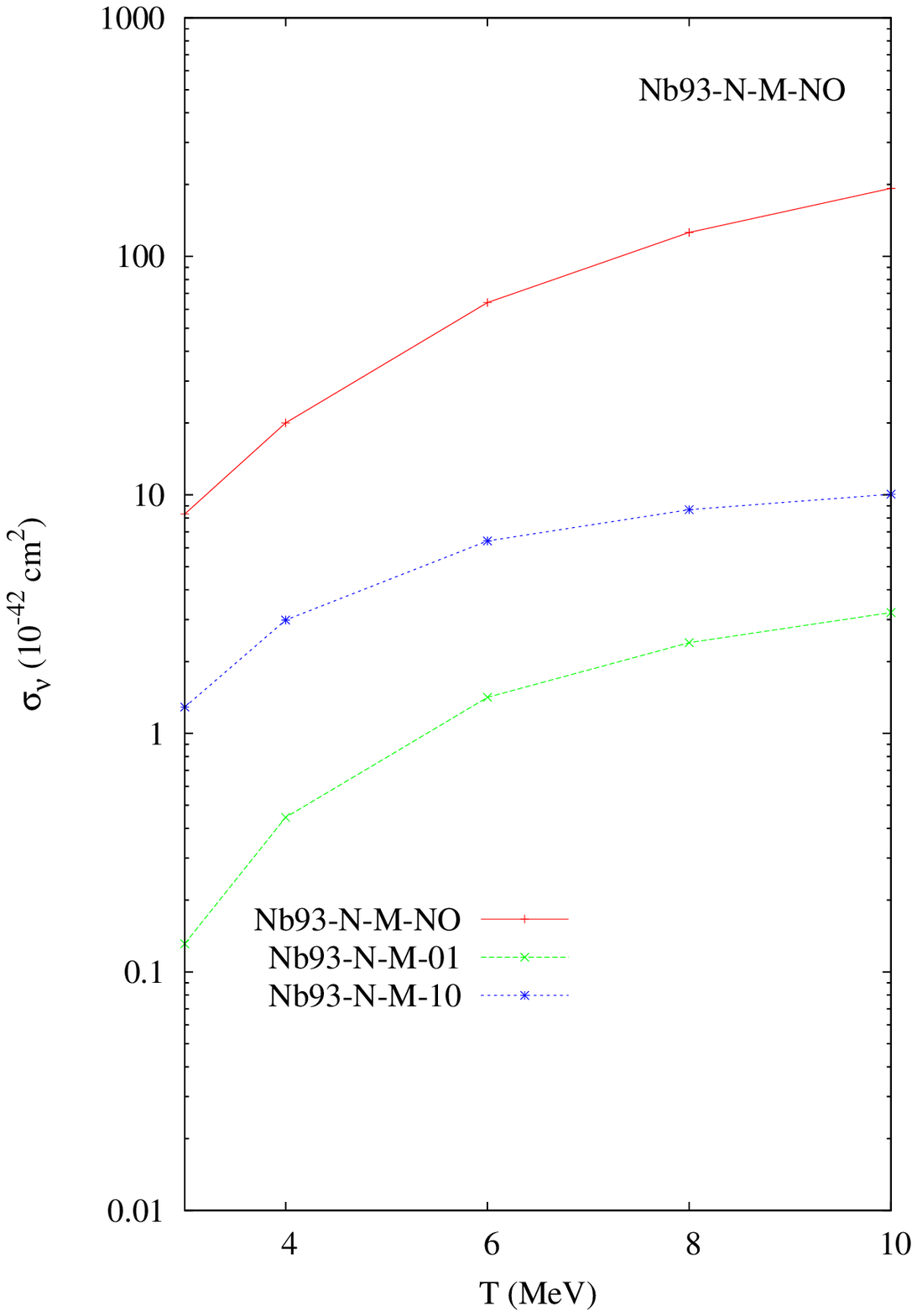}  \\
\includegraphics[width=7.2cm]{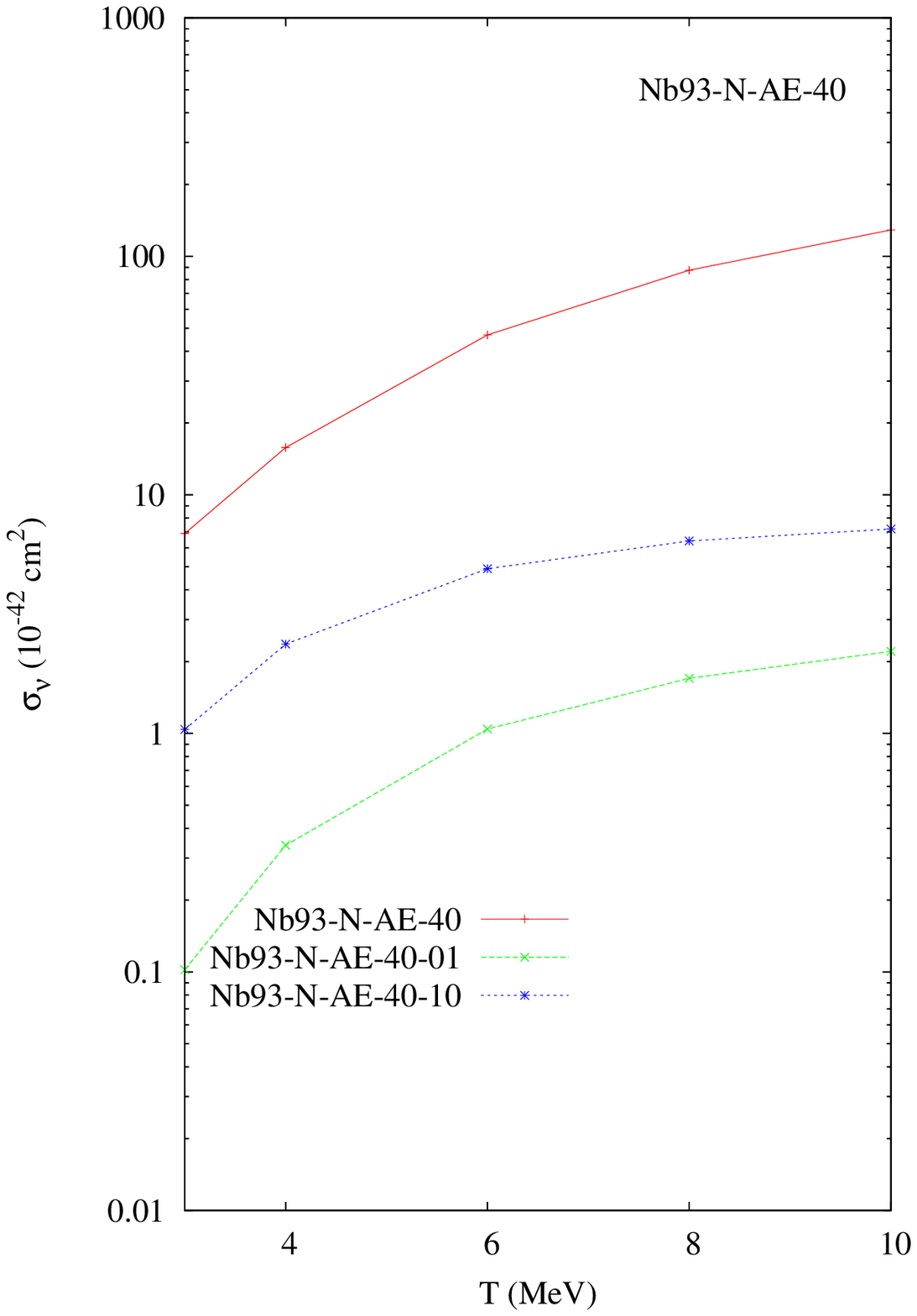}
\includegraphics[width=7.2cm]{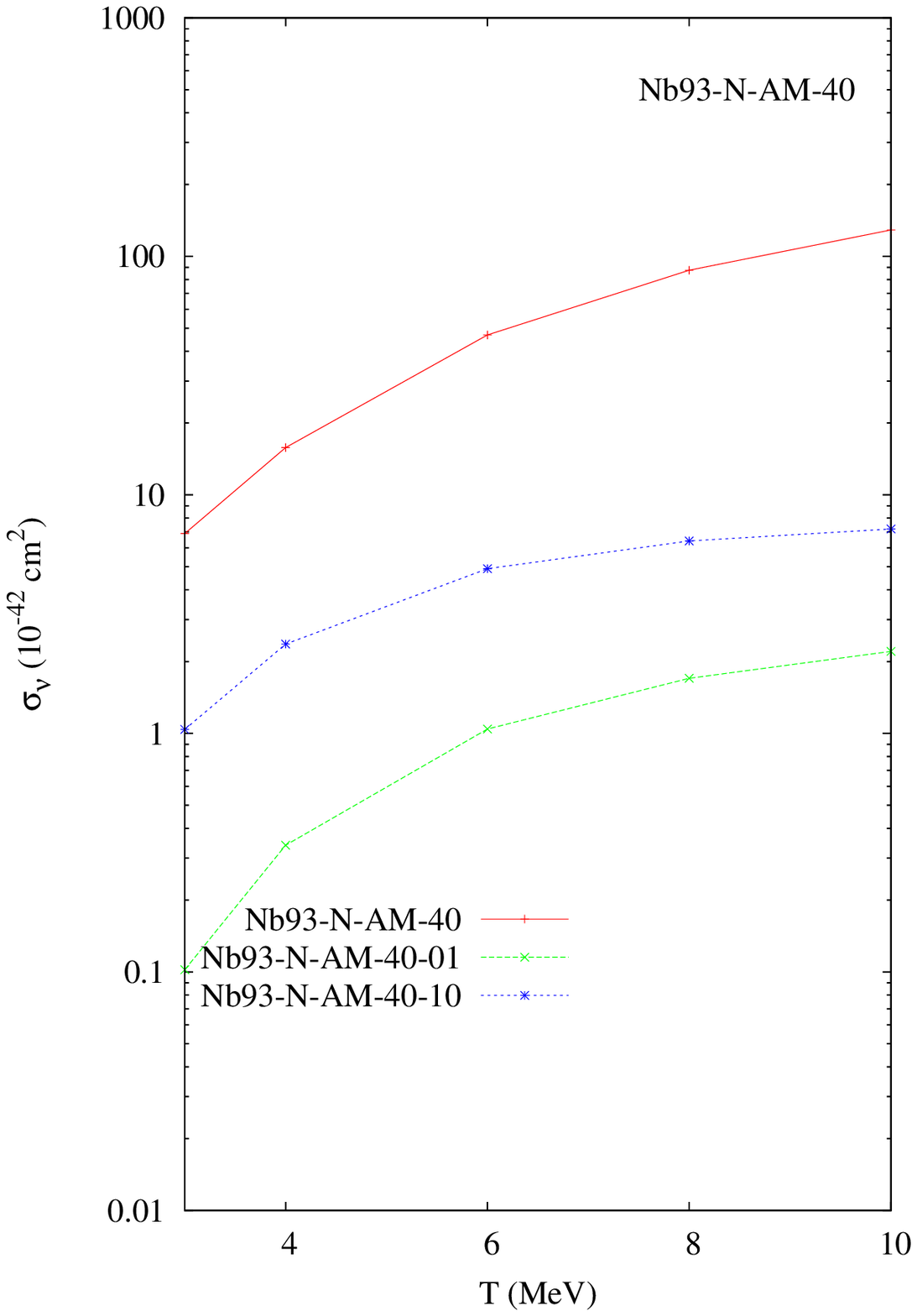}
\caption{Temperature dependent cross sections of NC reactions for
${}^{92}$Nb. The upper two figures are for incident $\nu_e$ (left) and
$\nu_\mu$ (right). The lower two results are for ${\bar \nu}_e$ (left)
and ${\bar \nu}_\mu$ (right). Red curves are ${}^{93}$Nb$( \nu
({\bar \nu}), \nu^{'} ({\bar \nu^{'}}) )$${}^{93}$Nb. Blue and
green curves are for proton and neutron emission decays from
${}^{93}$Nb, {\it i.e.} ${}^{93}$Nb$( \nu ({\bar \nu}), \nu^{'} ({\bar
\nu^{'}}) p)$${}^{92}$Zr and $^{93}$Nb$( \nu ({\bar \nu}), \nu^{'}
({\bar \nu^{'}}) n)$${}^{92}$Nb.} \label{fig.8}
\end{figure}

\end{document}